\documentclass[11pt,tightenlines,eqsecnum,floats,aps,amsmath,amssymb,nofootinbib,prd,floatfix]{revtex4}

\usepackage{graphicx, wrapfig}
\usepackage{array}
\usepackage{amsmath}
\usepackage{amsfonts}
\usepackage{amssymb}
\usepackage{empheq}
\usepackage{color}
\usepackage{mathrsfs}
\usepackage{subfigure}
\usepackage{mathrsfs}
\usepackage{subfigure}
\def\be{\begin{equation}}
\def\ee{\end{equation}}
\def\ba{\begin{eqnarray}}
\def\ea{\end{eqnarray}}

\newcommand{\pb}[1]{\hbox{\lower0.5ex\hbox{${}_{\leftarrow}$}}\kern-1.9ex{#1}}

\def\={\,\hat{=}\,}
\def\ord{\mathcal{O}}

\def\TT{\rm TT}
\def\tt{\rm tt}
\def\t{\rm t}
\def\T{\rm T}
\def\L{\rm L}
\def\Ac{{\underbar{A}}}
\def\phic{{\underline{\phi}}}

\def\1{(1)}
\def\2{(2)}
\def\bh{\bar{h}}

\def\ut#1{\rlap{\lower1ex\hbox{$\sim$}}{#1}}
\newcommand{\del}{\partial}
\newcommand{\grad}{\nabla}

\def\f{\frac}
\def\rmd{{\textrm{d}}}
\def\ub{\underbar}

\def\ti{\tilde}
\def\h{\hat}

\def\vk{\vec k}
\def\vx{\vec{x}}

\def\qo{\mathring{q}}

\def\Do{\mathring{D}}

\def\vx{\vec{x}}

\def\vA{\vec{A}}
\def\vh{\vec{h}} 
\def\hb{\bar{h}}

\def\scri{\mathcal{I}}
\def\scrip{\mathcal{I}^{+}}

\newcommand{\Lie}{\mathcal{L}}

\renewcommand{\d}{\mathrm{d}}

\newcommand{\re}{\text{Re}\,}
\newcommand{\im}{\text{Im}\,}

\begin{document}

\title{%Gravitational Waves in the Linearized Approximation:\\
%On the Conceptual Confusion in the Notion of Transverse-Traceless Modes
%On a basic conceptual confusion in gravitational radiation theory
On the ambiguity in the notion of transverse traceless modes\\ of gravitational waves}
\author{Abhay Ashtekar}
\email{ashtekar@gravity.psu.edu} \affiliation{Institute for
Gravitation and the Cosmos \& Physics
  Department, Penn State, University Park, PA 16802, U.S.A.}
\author{B\'eatrice Bonga}
\email{bpb165@psu.edu} \affiliation{Institute for Gravitation and
the Cosmos \& Physics
  Department, Penn State, University Park, PA 16802, U.S.A.}
%\author{Aruna Kesavan}
%\email{aok5232@psu.edu} \affiliation{Institute for Gravitation and the
%Cosmos \& Physics Department, Penn State, University Park, PA 16802,
%U.S.A.}

\begin{abstract}
Somewhat surprisingly, in many of the widely used monographs and review articles the term \emph{Transverse-Traceless modes} of linearized gravitational waves is used to denote two entirely different notions. These treatments generally begin with a decomposition of the metric perturbation that is \emph{local in the momentum space} (and hence non-local in  physical space), and denote the resulting transverse traceless modes by $h_{ab}^{\TT}$. However, while discussing gravitational waves emitted by an isolated system --typically in a later section-- the relevant modes are extracted using a `projection operator' that is \emph{local in physical space}. These modes are also called transverse-traceless and again labeled $h_{ab}^{\TT}$, implying that this is just a reformulation of the previous notion. But the two notions are conceptually distinct and the difference persists even in the asymptotic region. We show that this confusion arises already in Maxwell theory that is often discussed as a prelude to the gravitational case. Finally, we discuss why the distinction has nonetheless remained largely unnoticed, and also point out that there are some important physical effects where only one of the notions gives the correct answer.
\end{abstract}


\maketitle

\section{Introduction}
\label{s1}

In 1916 Einstein discovered that general relativity admits gravitational waves in the linearized approximation \cite{Einstein:1916}. Almost exactly a century later the LIGO collaboration announced the first direct detection of gravitational waves produced by coalescing black holes, thereby ushering-in the new field of gravitational wave astronomy \cite{ligo}. There are also ongoing missions to observe primordial gravitational waves \cite{primordial}. Physically, while waves that LIGO detects are produced by astrophysical sources, the origin of primordial radiation is cosmological. Mathematically, in the currently used theoretical paradigm, what LIGO observes is described by retarded solutions of Einstein's equations \emph{sourced by} highly dynamical compact objects in asymptotically flat space-times. What cosmological missions hope to observe is described by \emph{source-free} solutions of linearized Einstein's equations on a Friedmann-Lema\^{i}tre-Robertson-Walker background. Thus, not only are their observational techniques very different but the theoretical paradigms that underlie the two missions are also quite different. In particular, they use conceptually distinct notions of \emph{transverse-traceless} modes.

Unfortunately, much of the current literature on gravitational waves in general relativity, including some of the commonly used advanced texts as well as review articles, suggests that the two notions are the same and use the same symbol, $h^{\rm TT}_{ab}$, to denote them both. Our reading of several standard references and subsequent discussions led us to conclude that the conceptual confusion is rather widespread.  The goal of this work is to clarify the situation by spelling out what the two notions are, make it explicit that they are distinct, and discuss the relation between them in situations when both notions are available. The main issue arises in linearized gravity and the confusion we referred to is related to gravitational waves produced by isolated bodies. Therefore, in the main body of this paper we will restrict ourselves to this context and return to cosmological perturbations only at the end in section \ref{s4}. Thus, this article is primarily addressed to the community interested in gravitational waves produced by isolated bodies. 

Consider then linearized gravitational waves in Minkowski space-time. One introduces a $t=$\,{\rm const}  foliation by space-like planes $M_{t}$, and decomposes the space-time metric perturbation $h_{ab}$ into its irreducible parts. Of immediate interest is the decomposition of the spatially projected perturbation $\vh_{ab}$.  If  $\qo_{ab}$ denotes the flat, positive definite 3-metric on the $M_{t}$ slices, and $\Do$ its torsion-free connection, then we have the following  decomposition of $\vh_{ab}$ into its irreducible parts:
\be
\vh_{ab} = \f{1}{3} \qo_{ab} \, \qo^{cd} \vh_{cd} + \left(\Do_a \Do_b - \f{1}{3} \qo_{ab} \Do^2 \right) S + 2 \Do_{(a} \vec{V}_{b)}^{T} + h_{ab}^{\,\TT},  \label{decomposition}\ee
where $S$ is a scalar field, $\vec{V}_{a}^{\T}$ a transverse (spatial) vector field and $h_{ab}^{\,\TT}$ a symmetric, transverse-traceless tensor field:
\be \label{TT}
\Do^a \vec{V}_a^{\T} = 0 \qquad \Do^a h_{ab}^{\,\TT} =0 \qquad \qo^{ab} h_{ab}^{\,\TT} =0\, .
\ee 
This notion of $h_{ab}^{\TT}$ is widely used in several areas of physics. For instance, this is the decomposition of the metric perturbation used in standard  cosmology (where one is also interested in the scalar and vector modes). In quantum field theory in Minkowski space-time, one generally considers \emph{source-free} solutions to linearized Einstein's equations. Then the gauge invariant information in $h_{ab}$ is contained entirely in $h_{ab}^{\TT}$. The vector space of these fields can be naturally endowed with a Poincar\'e invariant Hermitian inner product. This is the Hilbert space of states of a graviton. It provides  irreducible representations of the Poincar\'e group with mass zero and helicity $\pm 2$. 

To extract $h_{ab}^{\TT}$ from $h_{ab}$ one typically goes to the momentum space where the operation is algebraic and hence local. See, e.g. Box 5.7 in \cite{pw}, or section 4.3 in \cite{straumann}, or section 35.4 of \cite{mtw}. By contrast the operation is non-local in physical space since it involves the inverse power of the Laplacian $\Do^{2}$. Thus, if one knows $h_{ab}$ only in a sub-region of the spatial manifold $M_{t}$ --say the asymptotic region-- one \emph{cannot determine $h_{ab}^{\TT}$ even in that sub-region.} Nonetheless, this notion of transverse-traceless modes is widely used because the field $h_{ab}^{\TT}$ is \emph{gauge invariant}.

But in the study of retarded fields produced by compact sources, most monographs and review articles switch to \emph{an entirely different} 
notion of transverse traceless modes which is \emph{local in physical space}. Specifically, one introduces a projection operator $P_{a}{}^{b}$ into the 2-sphere orthogonal to the radial direction in \emph{physical space} and extracts a \emph{new} transverse-traceless part of $h_{ab}$ by projecting $h_{ab}$ into the 2-sphere and removing the trace. This is again denoted by $h_{ab}^{\TT}$ implying that the projection operator $P_{a}{}^{b}$ provides just another way to extract what was previously called the transverse-traceless part of $h_{ab}$. See, e.g., chapter 11 of \cite{pw}, or section 4.5.1 in \cite{straumann}, or section 36.10 in \cite{mtw}, or section 1 of \cite{hughes}. All subsequent discussion of gravitational waves produced by isolated systems uses the transverse-traceless part of $h_{ab}$ that is extracted using $P_{a}{}^{b}$. Consequently, one sees only asymptotic expansions (in powers of $1/r$) of $h_{ab}$ in physical space;  Fourier transforms and/or inverse powers of Laplacians are completely absent in actual calculations of radiative modes, wave forms, and expressions of energy carried by gravitational waves.

This is confusing because the two notions of transverse-traceless parts are conceptually distinct and inequivalent. Therefore, let us change notation and set % change this notation --> change notation (for version 2 on arXiv)
\be \label{tt} \big( P_{a}{}^{c} P_{b}{}^{d} - \f{1}{2} P_{ab} P^{cd}\big)h_{cd}  =: h_{ab}^{\tt}\, . \ee
As we emphasized, the operation of extracting $h_{ab}^{\rm TT}$ from $h_{ab}$ is highly non-local in physical space and the resulting $h_{ab}^{\TT}$ is gauge invariant everywhere in space-time. On the other hand, the operation of extracting $h_{ab}^{\tt}$ is local in physical space and is not gauge invariant. However, in practice $h_{ab}^{\tt}$ is constructed only in the asymptotic region and its $1/r$-part can be shown to be gauge invariant under a large class of gauge transformations %\cite{pw} 
(although generally it is not explained why it suffices to restrict oneself to this class). As we will see, this second notion $h_{ab}^{\tt}$ is tailored to Bondi-Sachs type expansions \cite{bondi-sachs} and behavior of fields near null infinity. On the other hand, in the cosmological context, these Bondi-Sachs type expansions are not available and only the $h_{ab}^{\TT}$ notion is meaningful.

Still, in the asymptotically flat context, we have these two distinct notions of what we mean by transverse-traceless modes. What is the relation between them? Each notion leads to well-defined leading order asymptotic fields. Do these fields carry the same physical information? For example, can one construct expressions of energy, momentum and angular momentum carried by gravitational waves using either notion? What about gravitational memory, and `soft charges' that label the infrared sectors in the quantum theory? If one can use either methods to compute these physical quantities, how are the resulting expressions related? The purpose of this article is to address these issues using structure available at null infinity. We will show that, although the two notions are completely unrelated to start with, there is a large overlap in the physics they capture in the asymptotic region. However, there are also some important differences. \\

The paper is organized as follows. Since the central conceptual issues arise also for Maxwell fields in Minkowski space-time, in section \ref{s2} we discuss them in this technically simpler context. Again, there are two notions of transverse vector potentials, $A_{a}^{\T}$ and $A_{a}^{\rm \t}$. The first is gauge invariant but non-local in physical space, while the second is local in space but gauge invariant only asymptotically at $\scrip$ under suitably restricted gauge transformations. We will show that components of $A_{a}^{t}$ provide two functions which can be specified freely at $\scrip$, while components of $A_{a}^{\T}$ provide two freely specifiable functions \emph{and} a non-dynamical function (i.e., a function only of angles on $\scrip$). Thus, even at $\scrip$ there is more information in $A_{a}^{\T}$ than there is in $A_{a}^{\t}$. We will show that \emph{in presence of sources}, angular momentum carried away by electromagnetic waves and the total electric charge is \emph{not} expressible using just the two components of $A_{a}^{\t}$ but \emph{can be} expressed using the additional information carried by $A_{a}^{\T}$. In section \ref{s3} we discuss linearized gravitational fields, where the overall situation parallels that in the Maxwell case: in presence of sources, the wave forms of $h_{ab}^{\TT}$ and $h_{ab}^{\tt}$ are in general different even at $\scrip$ but the difference is time-independent. Again, the $1/r$-part of $h_{ab}^{\TT}$ contains `Coulombic' information that is not captured in the $1/r$-part of $h_{ab}^{\tt}$.%
\footnote{As discussed in section \ref{s3.3.4}, since there is no gauge invariant, local analog of a stress energy tensor for the linearized gravitational field, one has to introduce new structures to extract the full implications of this difference e.g. by calculating angular momentum carried by gravitational waves. This task is well beyond the scope of this work and will be undertaken elsewhere.}  
In section \ref{s4} we summarize the results and present the outlook.
%clarify some issues, such as why, although non-local in space, $h_{ab}^{\TT}$ is still relevant to gravitational wave detectors.

We use the following conventions. Throughout we assume that the underlying space-time is 4-dimensional and set $c$=1. The space-time metric has signature -,+,+,+.  The curvature tensors are defined via:  $2\nabla_{[a}\nabla_{b]} k_c = R_{abc}{}^d k_d$, $R_{ac} = R_{abc}{}^b$ and $R = R_{ab}g^{ab}$. Except when we are explicitly referring to components of tensor fields, we use Penrose's abstract index notation \cite{rpwr,ahm}. To avoid confusion, we will refer to $h_{ab}^{\rm TT}$ as `Transverse-Traceless' modes, and $h_{ab}^{\rm tt}$ as  `transverse-traceless' projection of $h_{ab}$. The main results of this paper were summarized in \cite{aabb1}. 

After the bulk of this work was completed, Badri Krishnan drew our attention to a pre-print by Istv\'{a}n R\'{a}cz \cite{racz1} in which he had already pointed out that the two notions of `transverse-traceless modes' used in the literature differ even asymptotically, and, after our pre-print \cite{aabb1} appeared on the arXiv, Istv\'{a}n R\'{a}cz brought to our attention another paper \cite{racz2} in which the electromagnetic case is discussed. Although the points of departure in these works are the same as in this paper, the goal and the main results are different. We will discuss the relation with that work at appropriate points.

\section{Maxwell Theory}
\label{s2}

Analysis simplifies for a Maxwell field because one can pass easily between Minkowski space-time and its conformal completion that includes null infinity, $\scri$, using the fact that the theory is conformally invariant. Therefore we will now discuss the main conceptual issues in this technically simpler context in some detail, and in the next section focus on points where gravity differs from Maxwell theory.
 
This section is divided into three parts. In the first, we use conformal invariance to show that for systems under consideration --where sources are supported on a spatially compact world tube-- the Maxwell field $F_{ab}$ satisfies the so-called `Peeling properties' at null infinity.
This discussion will also serve to fix notation and introduce future null infinity, $\scrip$. In the second part, we turn to vector potentials $A_{a}$ since in the gravitational case we are primarily interested in the properties of the linearized metric, not just its curvature. We obtain the minimal fall-off conditions that the potential must satisfy in virtue of field equations as one recedes from sources in null directions. Peeling properties of the Maxwell field simplify this discussion considerably. In the third part we introduce the two notions $A_{a}^{\T}$ and $A_{a}^{\t}$ of transversality and compare them. %Although  $A_{a}^{\T}$ and $A_{a}^{\t}$ differ even in the leading order (i.e., $\mathcal{O}(1/r)$) near $\scrip$, we show that the difference is encoded in non-dynamical fields. 
In particular we show that the presence of sources introduces an unforeseen twist: $A_{a}^{\T}$ carries certain physically important `Coulombic information' that escapes $A_{a}^{\t}$. 
%We then discuss energy-momentum and angular momentum carried by the electromagnetic waves. In the source-free case these quantities can be expressed in terms of the two radiative modes (and the same is true for gravitational waves -- even in full non-linear general relativity)  \cite{aams}. However, in presence of sources \emph{this is no longer the case for angular momentum;} its expression also involves the `Coulombic information' contained in $A_{a}^{\T}$ that escapes $A_{a}^{\t}$.
%In the fourth and the last part we summarize results and highlight the structural differences between the source-free case that the one with sources that is of more direct interest especially to the gravitational case.

\subsection{Null infinity and the Peeling behavior}
\label{s2.1}

As we will see in subsequent sub-sections, presence of sources introduces new features in the asymptotic properties and the physical content of Maxwell  potentials, even when the sources are confined to a spatially compact world tube. In that discussion we will assume certain asymptotic behavior of potentials. It is simplest to arrive at these fall-off properties by first noting the immediate consequences of the expressions of retarded solutions and then supplementing them with the implications of the Peeling properties of the Maxwell field. In the original discussion of Peeling \cite{np,etn}, it was assumed that a certain component ($\Phi_{0}$; see below) of the Maxwell field falls-off as $1/r^{3}$ and then the fall-off of other components ($\Phi_{1}$ and $\Phi_{2}$) was derived. Since it not a priori clear that the initial assumption is satisfied by retarded fields under consideration, for completeness we will now show that the Peeling properties hold also in presence of sources. This point is probably obvious to experts. We chose to include this discussion because the role of Peeling has been a focus of some of the recent discussions on soft gravitons and photons.%
\footnote{In the Christodoulou-Klainerman approach to the non-linear stability of Minkowski space-time \cite{dcsk}, the Weyl tensor components $\Psi_{1}$ and $\Psi_{0}$ do not peel in the standard manner, given, e.g., in \cite{np}. This feature is sometimes used to argue that failure of the standard Peeling behavior plays a crucial role in some of the discussion of infrared charges and soft gravitons and photons \cite{asgr21}. However, whether standard Peeling holds depends on the boundary conditions that the initial data satisfy. For example, it does hold in the Chrusciel-Delay \cite{pced} approach to non-linear stability of Minkowski space-time.  Infrared (or soft) photon charges as well as the new features we discuss arise also with standard Peeling shown in this sub-section.}
Our explicit demonstration will make it clear that new features discussed in this paper arise even with the standard Peeling behavior.

\subsubsection{Future null infinity $\scrip$}
\label{s2.1.1}

Let us begin with a concrete conformal completion of Minkowski space $(M, \eta_{ab})$ that focuses on $\scrip$. In terms of the retarded spherical coordinates $u=t-r, r, \theta, \varphi$, we have
\be \rmd s^{2} = \eta_{ab} \rmd x^{a}\rmd x^{b} = -\rmd u^{2} - 2\rmd u  \rmd r + r^{2}\,(\rmd \theta^{2} + \sin^{2}\theta\, \rmd\varphi^{2})\, . \ee
Let us conformally rescale $\eta_{ab}$ with a smooth conformal factor $\Omega$ with $\Omega =\f{1}{r}$ outside the world-tube $r=r_{0}$ for some $r_{0}$, and attach to $M$ a boundary $\scrip$ at which $\Omega$ vanishes. Then, for $r> r_{0}$ the conformally rescaled metric $\h\eta_{ab}$ is given by 
 \be \rmd \h{s}^{2} = \h\eta_{ab} \rmd x^{a}\rmd x^{b} = -\Omega^{2}\rmd u^{2} + 2\rmd u  \rmd \Omega + \,(\rmd \theta^{2} + \sin^{2}\theta\, \rmd\varphi^{2})\, . \ee
The rescaled metric $\h\eta_{ab}$ is well-defined everywhere on the manifold with boundary $\h{M} = M\cup \scrip$. Since $\Omega$ vanishes on the boundary $\scrip$, it is coordinatized by $u \in (-\infty,\infty)$ and $\theta,\varphi \in \mathbb{S}^{2}$. It is thus topologically $\mathbb{S}^{2}\times \mathbb{R}$,\, with a null normal\,\, $\ti{n}^{a} \= \h\eta^{ab}\nabla_{b}\Omega$\,\, that satisfies $\h\nabla_{a}\ti{n}_{b} \= 0$, where from now onwards \emph{$\=$ stands for equality restricted to the points of the boundary $\scrip$.} In particular, the conformal frame is `divergence free'. Furthermore, the pull-back of $\h\eta_{ab}$ to $\scrip$ is the unit 2-sphere metric $q_{ab}$ with $q_{ab}\, \rmd x^{a} \rmd x^{b} = \rmd\theta^{2} + \sin^{2}\theta\,\, \rmd\varphi^{2}$. Thus, with this conformal completion we are in a Bondi conformal frame at $\scrip$ (see, e.g., \cite{aa-bib}). A general Killing field $K^{a}$ of $\eta_{ab}$ has the limit
\be \label{killing}  K^{a}\, \= \, \big[\alpha(\theta,\varphi) + u \beta(\theta,\varphi)\big]\, \ti{n}^{a} +  h^{a}  \ee
to $\scrip$. Here $\alpha(\theta,\varphi)$ is linear combinations of the first four spherical harmonics, $Y_{00}$ and $Y_{1m}$;\, $\beta(\theta,\varphi)$ of the three $Y_{1,m}$;\, and $h^{a}$ is a `horizontal' (i.e., tangential to the $u \={\rm const}$ 2-sphere cross-sections of $\scrip$) and conformal Killing field of the unit 2-sphere metric $S_{ab}^{0}$ thereon, $\mathcal{L}_{h} S_{ab}^{0} \= 2\beta S_{ab}^{0}$,\, satisfying  $\mathcal{L}_{h} \ti{n}^{a} \=0$ (see, e.g., \cite{tdms,ak-thesis}).
In particular, the translation Killing fields of $\eta_{ab}$ are represented by $\alpha \ti{n}^{a}$. We will use these facts in section \ref{s2.3}. 

\subsubsection{Peeling}
\label{s2.1.2}
Let us suppose we are given a 4-current $j^{a}(t,\vx)$ which is smooth and of compact spatial support, and whose Cartesian components remain uniformly bounded in time. Consider the retarded solution of Maxwell's equations on Minkowski space-time $(M,\, \eta_{ab})$ with $j^{a}$ as source:
\be \label{max1}  \rmd\, {F} =0, \quad {\rm and}\quad \nabla^{a}F_{ab} = -4\pi j_{b}\quad {\Leftrightarrow} \quad \rmd\, {}^{\star}\!{F} = 4\pi\,\,{}^{\star}\!{j}, \ee
where ${}^{\star}\!j_{abc} = \epsilon_{dabc} j^{d}$. The question is whether the limit to $\scrip$  --i.e., the limit $r \to \infty$ keeping $u, \theta,\varphi$ constant-- of the retarded solution $F_{ab}$ to these equations satisfies the standard Peeling properties.

To answer this question, one can directly calculate the retarded solution $F_{ab}$ and examine its asymptotic behavior in detail. However, it is much simpler to use conformal invariance of Maxwell's equations. By inspection it follows that $\h{F}_{ab} := F_{ab}$ and ${}^{\star}\!\h{F}_{ab} = \f{1}{2}\h{\epsilon}_{ab}{}^{cd}\! \h{F}_{cd} = {}^{\star}F_{ab}$ also satisfy Maxwell's equations on $(\hat{M}, \h\eta_{ab})$ :
\be \rmd\, \h{F} =0, \quad {\rm and} \quad \rmd {}^{\star}\,\h{F} = 4\pi\,\,{}^{\star}\!\h{j}\, , \ee
with ${}^{\star}\!\h{j} = {}^{\star}{j}$. Since $\scrip$ is just a sub-manifold of the conformally completed space-time $(\h{M},\, \h{\eta}_{ab})$ that is `a finite distance away from sources', it follows that $\h{F}_{ab}$ is smooth on $\scrip$. We will now show that this fact implies that $F_{ab}$ satisfies the standard Peeling properties.

For this, let us use the Newman-Penrose null co-tetrad on Minkowski space-time $(M,\,\eta_{ab})$, given by%
\ba n_{a} &=& -\f{1}{\sqrt{2}}\, (\nabla_{a} t + \nabla_{a} r),\quad \ell_{a} = - \f{1}{\sqrt{2}}\, (\nabla_{a} t - \nabla_{a} r),\nonumber\\ m_{a} &=& \f{r}{\sqrt{2} }\, (\nabla_{a} \theta + {i}\,{\sin\theta}\, \nabla_{a}\phi), \quad  \bar{m}_{a} = \f{r}{\sqrt{2}}\,(\nabla_{a} \theta - {i}\, {\sin\theta}\, \nabla_{a}\phi)\, .\ea
The Newman-Penrose null tetrad is obtained simply by raising indices of these 1-forms with $\eta^{ab}$. It is straightforward to check that  
\be \label{hatted} \h{n}^{a} := n^{a},\,\, \h{\ell}^{a} := r^{2}\,\ell^{a},\,\, \h{m}^{a} := r\, m^{a},\,\,{\rm and}\,\, \h{\bar{m}}^{a} := r\,\bar{m}^{a} \ee
have smooth, non-vanishing limits to $\scrip$ and define a null tetrad there.% 
\footnote{\label{ntilde} Note that, with these conventions, at $\scrip$ we have $\ti{n}^{a}\, \hat{=}\, \f{1}{\sqrt{2}} \h{n}^{a}$, and $\ti{n}^{a}$ is the limit to $\scrip$ of the unit time translation $t^{a} = -\eta^{ab}\nabla_{b}t$ of the Minkowski metric $\eta_{ab}$.} 
Therefore, components of the Maxwell fields $\h{F}_{ab}$ in the hatted null tetrad have smooth limits to $\scrip$. This in turn implies that components of $F_{ab}$ in the Newman-Penrose tetrad in Minkowski space-time have the following asymptotic behavior
\ba 
\Phi_2 &:=& F_{ab} n^a \bar{m}^b = \f{1}{r}\,\, \h{F}_{ab} \h{n}^a \h{\bar{m}}^b = \f{\Phi_{2}^{0}(u,\theta,\varphi)}{r} + \mathcal{O}\Big(\f{1}{r^{2}}\Big), \label{peeling1}\\
\Phi_1 &:=& \frac{1}{2} F_{ab} \left( n^a l^b + m^a \bar{m}^b \right) =  
\f{1}{2r^{2}}\,\, \h{F}_{ab} \left( \h{n}^a \h{l}^b + \h{m}^a \h{\bar{m}}^b \right) = \f{\Phi_{1}^{0}(u,\theta,\varphi)}{r^{2}} + \mathcal{O}\Big(\f{1}{r^{3}}\Big), \label{peeling2}\\
\Phi_0 &:=& F_{ab} m^a l^b = \f{1}{r^{3}}\,\, \h{F}_{ab} \h{m}^a \h{l}^b = \f{\Phi_{0}^{0}(u,\theta,\varphi)}{r^{3}} + \mathcal{O}\Big(\f{1}{r^{4}}\Big)\, ,
\label{peeling3}\ea
since the hatted fields have well-defined limits as $r\to \infty$, keeping $u,\theta,\varphi$ constant. These are precisely the standard Peeling properties of the Maxwell field \cite{np}. Presence of sources under consideration does not introduce any new element in this result. Fields $\Phi_{2}^{0}, \, \Phi_{1}^{0}$ and $\Phi_{0}^{0}$ on $\scrip$ encode the leading order, asymptotic Maxwell field. Maxwell's equations leave $\Phi_{2}^{0}$ unconstrained. Because of this property, and because $\Phi_{2}$ falls off as $1/r$, $\Phi_{2}^{0}$ is called the `radiation field' on $\scrip$. The time derivatives of $\Phi_{1}^{0}$ and $\Phi_{0}^{0}$ are determined by angular derivatives of $\Phi_{2}^{0}$: %
\footnote{\label{eth}Fields such as $\Phi_{2}, \Phi_{1}$ and $\Phi_{0}$ depend on the choice of the dyad $\h{m}^{a}, \hat{\bar{m}}^{a}$ on the $(\theta-\varphi)$ 2-sphere. A field $f$ is said to have spin weight $s$ if it transforms as $f \to e^{is\chi}f$ under the dyad rotation $\h{m}^{a} \to e^{i\chi}\h{m}^{a}$. The angular derivative $\eth$ of a spin $s$ weighted field is defined by $\eth f = \f{1}{\sqrt{2}}\,(\h{m}^{a}D_{a} f  - \f{s}{\sqrt{2}}\, \cot\theta\, f) \equiv \f{1}{2}\, (\partial_{\theta}f + \f{i}{\sin\theta}\partial_{\phi} f - s \cot\theta f)$, where $D$ is the derivative operator on a unit 2-sphere. Similarly, $\bar{\eth} f =   \f{1}{2}\, (\partial_{\theta}f - \f{i}{\sin\theta}\partial_{\phi} f + s \cot\theta f)$.}
\be \label{phieqs} \partial_{u}\Phi_{1}^{0} = \eth \Phi_{2}^{0}, \qquad {\rm and} \qquad  \partial_{u}\Phi_{0}^{0} = \eth \Phi_{1}^{0}\, .\ee
In this sense $\Phi_{1}^{0}$ and $\Phi_{0}^{0}$ do not have any dynamical degrees of freedom of their own: given $\Phi_{2}^{0}$, they  are determined by their values at spatial infinity $i^{o}$ (i.e., $u=-\infty$), or, time-like infinity $i^{+}$ (i.e. $u=\infty$.).

In the next sub-section we will use these fall-off properties (\ref{peeling3}) to find useful relations between components of the vector potential that hold in a broad class of gauges.

\subsection{Asymptotic conditions on potentials}  
\label{s2.2}

We will need asymptotic conditions that are satisfied by vector potentials $A_{a}$ of our retarded Maxwell fields. For these potentials we cannot just invoke conformal invariance as we did for $F_{ab}$ because the gauge conditions of interest generally fail to be conformally invariant. A seemingly natural strategy is to assume that, since $F_{ab}$ admits a smooth limit to $\scrip$, one should simply require that the potential $A_{a}$ is also smooth there. However, in the Coulomb gauge that is used to define Transversality, $A_{a}$ fails to satisfy this condition even in the simple case when the source is a point charge (see below). Therefore, we need to first find an appropriate set of fall-off conditions for $A_{a}$. Given a slicing of Minkowski space by hyperplanes $t= {\rm const}$, we can decompose the 4-potential $A_{a}$ as follows:
\begin{align} 
A_a &= - \phi \grad_a t + \vec{A}_a  \nonumber\\
&= - \phi \grad_a u + \left(- \phi +  A_1 \right)\grad_a r + A_2\, m_a + \bar{A}_2\, \bar{m}_a. \label{eq:vecexp}
\end{align}
We will first examine the implications of field equations on the asymptotic behavior of $\phi, A_{1}$ and $A_{2}$ in the Coulomb gauge.   The arguments that follow are only meant to motivate the rather weak fall-off conditions we will use in the paper; they do not constitute a rigorous derivation.\\

Let us begin by stating the assumptions. We consider smooth vector potentials $A_{a}$ which vanish at spatial infinity for all $t$ and satisfy  $\Do^{a} \vA_{a} =0$. These conditions exhaust the initial gauge freedom $A_{a} \to A_{a} + \nabla_{a} \Lambda$.
%for all $t$, vanish as $r \to \infty$, and satisfy  $\Do^{a} \vA_{a} =0$. Then the initial gauge freedom $A_{a} \to A_{a} + \nabla_{a} \Lambda$ reduces to $\vA_{a} \to \vA_{a}$ and  $\phi \to \phi - \partial_{t}\Lambda(t)$. We will exhaust it by demanding that $\phi \to 0$ as $r\to \infty$ for all $t$. 
To write down the field equations on $\phi$ and $\vA_{a}$, it is useful to first note a few consequences of our initial assumptions on the source current $j_{a}$ 
stated in section \ref{s2.1.2}. Since $j_{a}$ is smooth and of compact spatial support, on each $t={\rm const}$ slice its spatial projection $\vec{j}_a$ is in particular in the Schwartz space $\mathcal{S}$: Cartesian components of $\vec{j}_a$ are $C^{\infty}$ functions that, together with all its derivatives, fall off faster than the inverse of any polynomial in $r$ as $r\to \infty$ on $t={\rm const}$ surfaces. Now, $\mathcal{S}$ is stable under Fourier transform. Hence the Fourier transform of $\vec{j}_{a}$ is also in the Schwartz space in the momentum space. Because the operation of extracting the Transverse part is algebraic in momentum space, it follows immediately that $\vec{j}_{a}^{\,\T}(\vk)$ decays faster than any inverse polynomial in $|\vk|$ as $|\vk| \to \infty$. However, since the projection operator into the Transverse part projects $\vec{j}_{a}(\vk)$ into 2-spheres centered at the origin, $\vec{j}_{a}^{\,\T}(\vk)$ also fails to be smooth there. Therefore, although it is smooth everywhere else, bounded at the origin, and decays rapidly as $|\vk| \to \infty$, in general $\vec{j}_{a}^{\,\T}(\vk)$ is not in $\mathcal{S}$. So neither is $\vec{j}_{a}^{\,\T}(\vx)$ in the physical space in the Schwartz space $\mathcal{S}$. Nonetheless, properties of $\vec{j}_{a}^{\,\T}(\vk)$ we just summarized imply that the physical space $\vec{j}_{a}^{\,\T}(\vx)$ is smooth, with $\int  \vec{j}_{a}^{\,\T}(\vx)\, \rmd^{3} x$ bounded. %
%
%\footnote{In these considerations $\vec{j}_{a}^{\,\T}(\vx)$ is regarded as a distribution, integrated against test functions in $\mathcal{S}$. Its integral in space is obtained as a limit of a suitable sequence of integrals where the test functions tend to 1$.} 
%   
(In the terminology one often uses in physics literature, the boundedness of the integral can be taken to mean that the Cartesian components of $\vec{j}_{a}^{\,\T}(\vx)$ fall-off faster than $1/r^{3}$.) 

In particular, then, the decomposition of $j_{a}$ into Transverse and Longitudinal parts is well-defined on each $t={\rm const}$ surface: 
\begin{align*}
j_a = - \rho \, \grad_a t + \vec{j}_a\,\, 
= - \rho \, \grad_a t + \vec{j}_a^{\,\T} + \Do_a j^{\, \rm L} 
\end{align*}
with $\Do^a \vec{j}_a^{\,\T} =0$. Therefore we can write Maxwell's equations \emph{in the Coulomb gauge} as:
\begin{align}
\Do^2 \phi &= - 4 \pi \rho \quad {\rm and} \label{max1}\\
\Box \vec{A}_a &= - 4 \pi \left( \vec{j}_a - \frac{1}{4 \pi} \Do_a \dot{\phi} \right) = - 4 \pi \vec{j}_a^{\,\T}\, , \label{max2}
\end{align}
where in the second step we have used (\ref{max1}) and the conservation of 4-current. The equation for $\phi$ can be solved on each $t={\rm const}$ slice and we consider the retarded solution for $\vec{A}_{a}$:
\begin{align}
\phi(t,\vec{x}) &= \int \d ^3x' \,\, \frac{\rho(t,\vec{x}')}{\left| \vec{x} - \vec{x}' \right|}  \\
\vec{A}_a (t,\vec{x}) &= \int \d ^3x' \,\,  \frac{\vec{j}_a^{\,\T}(t-\left| \vec{x} - \vec{x}' \right|,\vec{x}')}{\left| \vec{x} - \vec{x}' \right|}\, . 
\end{align}
Recall that the matter current is smooth, of compact spatial support,  \emph{and} uniformly bounded in time. Therefore, an examination of the integral on the right shows that the solutions $\phi$ and (the Cartesian components of) $\vec{A}_{a}$ fall-off at least as fast as $\mathcal{O}(\f{1}{r})$ as $r\to \infty$ along constant $u, \theta, \varphi$ directions. Therefore in the Coulomb gauge we are led to the following asymptotic behavior: 
\begin{align}
\phi &= \frac{\phi^{0} (u, \theta,\varphi)}{r} + \frac{\phi^{1} (u, \theta,\varphi)}{r^2} + \ldots \label{falloff1}\\
A_1 &= \frac{A_1^0 (u, \theta,\varphi)}{r} + \frac{A_1^1 (u, \theta,\varphi)}{r^2} + \ldots \label{falloff2} \\
A_2 &= \frac{A_2^0 (u, \theta,\varphi)}{r} + \frac{A_2^1 (u, \theta,\varphi)}{r^2} + \ldots \label{falloff3}
\end{align}
where the coefficients $\phi^{0}, \ldots , A_{2}^{1}$ are smooth functions (of appropriate spin weights) on $r={\rm const}$ surfaces. Although we decomposed the 4-potential using a specific foliation by space-like hyperplanes, the asymptotic conditions (\ref{falloff1}) - (\ref{falloff3}) are insensitive to this choice.\\ 

\emph{Remark:}\, Although $\vec{j}_{a}$ is of compact support, the support of its Transverse part $\vec{j}_a^{\,\T}$ extends to spatial infinity (although, as we saw, $\vec{j}_a^{\,\T}$ falls-off sufficiently fast for its integral to be finite.) Frenkel and R\'{a}cz pointed out in \cite{racz2} that this fact introduces a number of complications that have been generally overlooked in the literature. The main goal of that work was to calculate $A_{a}^{\T}$ and $A_{a}^{\t}$ in specific examples and show that they differ even in the asymptotic region. In these examples, conditions (\ref{falloff1}) - (\ref{falloff3}) are satisfied both in the Coulomb and Lorenz gauges. However, as we show in section \ref{s3.3.3} one of  their conclusions, ``all the troubles yielded by the replacement of the proper projection operator by a simplified transversal one goes away once gauge-independent quantities are applied,'' is not borne out because `the simplified projection' $A_{a}^{\t}$ misses the `Coulombic information' in $F_{ab}$ and there is a subtle interplay between the radiative and Coulombic aspects.
\\

Using this result in the Coulomb gauge as motivation, \emph{from now on we will restrict ourselves to gauges in which the 4-potential has the fall-off behavior given above.} %Explicit examples involving point charges and dipoles show that these boundary conditions also hold in the Lorenz gauge and are thus physically reasonable (see also \cite{racz2}). 
Although these fall-off conditions are weak, field equations imply further restrictions on the coefficients. We will find that the sum total of these conditions is sufficient to arrive at physically interesting results. 

Next, let us examine the limit of the vector potential to $\scrip$. For this we have to express $\phi$ and $\vec{A}_{a}$ of (\ref{eq:vecexp}) in terms of basis vectors in the conformally completed space-time $(\h{M}, \h{\eta}_{ab})$ which are well-behaved at $\scrip$ (see \ref{hatted}) and footnote \ref{ntilde}). The result is:
\ba A_{a} &=& \sqrt{2} \phi \h\ell_{a} + \Omega^{-2}(\phi-A_{1}) \ti{n}_{a} + \Omega^{-1}A_{2} \h{m}_{a} + \Omega^{-1} \bar{A}_{2} \h{\bar{m}}_{a}\nonumber\\
&=& \sqrt{2} \big(\Omega\phi^{0} + \Omega^{2}\phi^{1} + \ldots\big)\h\ell_{a} \,+\, \big( \Omega^{-1}\phi^{0} + \phi^{1} + \ldots -\Omega^{-1} A_{1}^{0} - A_{1}^{1} - \ldots \big) \ti{n}_{a}\nonumber\\ 
&+&\, \big(A_{2}^{0} + \Omega A_{2}^{1}+ \ldots \big)\h{m}_{a}\, + \,
\big(\bar{A}_{2}^{0} + \Omega \bar{A}_{2}^{1}+ \ldots \big)\h{\bar{m}}_{a}\, . \label{asymexpn}
\ea
Thus, in spite of the fall-offs (\ref{falloff1}) - (\ref{falloff3}), because of the presence of $\Omega^{-1}$ terms the 4-potential $A_{a}$ \emph{diverges at $\scrip$} in the conformally completed space-time, unless $\phi$ and $A_{1}$ fall-off faster than $1/r$. However, as noted in the beginning of this section, the stronger fall-off conditions are not met even for a static point charge in Minkowski space. In the source-free context, one often requires that $A_{a}$ should admit a smooth limit at $\scrip$ and this requirement does not rule out interesting gauges. However, in presence of sources, this is no longer the case: now that requirement would not allow us to use the Transverse gauge! In the more general setting of (\ref{asymexpn}), the leading order asymptotic fields, $\phi^{0}, A_{1}^{0}$ and $A_{2}^{0}$ are now given by limits to $\scrip$ of $\Omega^{-1} A_{a}\h{n}^{a},\,\Omega A_{a}\h\ell^{a}$ and $A_{a}\h{\bar{m}}^{a}$ respectively. Finally, as we will see, the transverse field $A_{a}^{\t}$ at $\scrip$ knows only about $A_{2}^{0},\, \bar{A}_{2}^{0}$. This information can also be extracted from $A_{a}$ by first taking the pull-back $\pb{{A}_{a}}$ to the $\Omega={\rm const}$ surfaces of $A_{a}$ and then taking the limit to $\scrip$: $\lim_{\scrip} \pb{{A}_{a}} = A_{2}^{0}\h{m}_{a} + \bar{A}_{2}^{0} \h{\bar{m}}^{a}$. \\

%However, note that the divergent term is proportional to $\ti{n}_{a} = \nabla_{a}\Omega$. Therefore if we were to first pull-back the 3-potential $\vec{A}_{a}$ to the $\Omega={\rm const}$ surfaces to obtain
%
%\be \pb{\vec{A}_{a}} = \sqrt{2} \big(\Omega\phi^{0} + \Omega^{2}\phi^{1} + \ldots\big)\h\ell_{a} \, +\, \big(A_{2}^{0} + \Omega A_{2}^{1}+ \ldots \big)\h{m}_{a}\, + \, \big(\bar{A}_{2}^{0} + \Omega \bar{A}_{2}^{1}+ \ldots \big)\h{\bar{m}}_{a}\, ,\ee
%
%and then take the limit to $\scrip$, the limit would be smooth. The limit carries information about the leading order fields $A_{2}^{0},\, \bar{A}_{2}^{0}$ but not $A_{1}^{0}$. However, we will see that the Peeling properties of the Maxwell field imply that the information in $A_{1}^{0}$ is already contained in the scalar potential $\phi$. Therefore, while discussing asymptotic properties at $\scrip$, we will only assume that: (i) $\Omega^{-1}\phi$ and the pulled-back vector potential $\pb{\vec{A}_{a}}$ admit smooth limits to $\scrip$, encoded in $\phi^{0}$ and $A_{2}^{0}$, and, (ii) these limits have suitable fall off as $u\to \pm \infty$ along $\scrip$ to guarantee finiteness of energy, momentum and angular momentum. \emph{Thus, the notion of null infinity provides a succinct characterization of the class of Maxwell potentials of interest to our investigation.} \\

As noted above, Maxwell's equations --and, in particular, the Peeling properties they imply-- can be used to further restrict this asymptotic behavior of $A_{a}$ \emph{irrespective of the gauge choice} so long as it meets the requirements (\ref{falloff1}) - (\ref{falloff3}). Let us begin by expressing the Newman-Penrose components of the asymptotic Maxwell field in terms of the vector potential:
\begin{align}
\Phi_2 &= \frac{\sqrt{2}}{r} \del_u A_2^0 + \ord \Big(\f{1}{r^{2}}\Big) 
\equiv \f{\Phi_{2}^{0}}{r} + \ord \Big(\f{1}{r^{2}}\Big) \nonumber\\
%%+ \frac{1}{r^2} \left[ \eth (A_1^0 + \phi_0) + \sqrt{2} \del_u A_2^1 \right] + \ord(r^{-3}). 
%\label{eq:phi2}\\
%%%%%%%%%%%%%%%%%%%%%%%%%%%%%%%%%%%%%%%%%%%%%%%%%%%%%%%%%%%
\Phi_1 &= \frac{1}{2r}  \underbrace{\del_u(A_1^0 -\phi^{0})}_{\text{ =0  by (\ref{peeling2})}} + \frac{1}{2r^2} \left[ - \phi^{0} + \del_u (A_1^1 - \phi^{1}) +  \sqrt{2} \eth A_2^0 - \sqrt{2}\bar{\eth} \bar{A}_2^0 \right] + \ord \Big(\f{1}{r^{3}}\Big)  \equiv  \f{\Phi_{1}^{0}}{r^{2}} +
\ord \Big(\f{1}{r^{3}}\Big) 
 \nonumber\\
%\label{eq:phi1}\\
%%%%%%%%%%%%%%%%%%%%%%%%%%%%%%%%%%%%%%%%%%%%%%%%%%%%%%%%%%%
\Phi_0 &= \frac{1}{r^2} \, \underbrace{\eth( A_1^0 - \phi^{0})}_{\text{ =0 \, by (\ref{peeling3})}} + \frac{1}{r^3} \left[ \eth (A_1^1 - \phi^{1}) + \frac{1}{\sqrt{2}} \bar{A}_2^1 \right] +  \ord \Big(\f{1}{r^{4}}\Big) \equiv \f{\Phi_{0}^{0}}{r^{3}} + \ord \Big(\f{1}{r^{4}}\Big) \, .
\label{Phi} 
\end{align}
Here $\eth$ is again the Newman-Penrose \cite{np} angular derivative operator on spin-weighted functions (see footnote \ref{eth}).
%
%\footnote{Fields such as $\Phi_{2}, \Phi_{1}$ and $\Phi_{0}$ depend on the choice of the dyad $\h{m}^{a}, \hat{\bar{m}}_{a}$ on the $(\theta-\varphi)$ 2-sphere. A field $f$ is said to have spin weight $s$ if it transforms as $f \to e^{is\chi}f$ under the dyad rotation $\h{m}^{a} \to e^{i\chi}\h{m}^{a}$. The angular derivative $\eth$ of a spin $s$ weighted field is defined by $\eth f = \f{1}{\sqrt{2}}\,(\h{m}^{a}D_{a} f  - \f{s}{\sqrt{2}}\, \cot\theta\, f) \equiv \f{1}{2}\, (\partial_{\theta}f + \f{i}{\sin\theta}\partial_{\phi} f - s \cot\theta f)$, where $D$ is the derivative operator on a unit 2-sphere. }
%

These equations have several noteworthy features that will play an important role in the subsequent analysis.

\begin{itemize}
\item First, the asymptotic `radiation field' $\Phi_{2}^{0}$ is determined by the time derivative of $A_{2}^{0}$ in any choice of gauge: $\Phi_{2}^{0} = \sqrt{2} \partial_{u} A_{2}^{0}$. Therefore, the complex field $A_{2}^{0}$ at $\scrip$ represents the \emph{two radiative modes} of the Maxwell field at $\scrip$. Note that, through its angular derivatives, $A_{2}^{0}$ also determines $\im{\Phi_{1}^{0}}$.  

\item The fields $\re{\Phi_{1}^{0}}$ and $\Phi_{0}^{0}$ carry the additional `Coulombic information' in the Maxwell field at $\scrip$.  For example, the projection along $\h{n}^{b}$ of the source-free Maxwell equation $\h\nabla^{a}\h{F}_{ab} \=0$ reads \cite{np} 
\be \label{Phi1dot} \partial_{u}\, \re \Phi_{1}^{0} \= \re \big(\eth\, \Phi_{2}^{0}\big) \= \sqrt{2} \partial_{u}\,\re\big(\eth\, A_{2}^{0}\big), . \ee
From the first equality (which we had already noted in Eq. (\ref{phieqs})) it follows that 
\be \label{charge1} Q \= - \f{1}{2\pi} \oint \re\Phi_{1}^{0}(u,\theta,\varphi) \, \rmd^{2}S \ee 
is conserved, where the integral is taken over a 2-sphere cross-section of $\scrip$. $Q$ is of course the total electric charge of the source. The second equality in (\ref{Phi1dot}) implies 
\be \label{rephi10} \re\Phi_{1}^{0} \= \sqrt{2}\, \re \eth A_{2}^{0} + G(\theta, \varphi)\, .
\ee
$G(\theta,\varphi)$ is the `integration constant', which can also be expressed as $2G(\theta,\varphi) = -\phi^{0} + \partial_{u}(A_{1}^{1} - \phi^{1})$ in any gauge in our class (see the second  equation in (\ref{Phi})). \emph{It is this non-dynamical, real function $G(\theta,\varphi)$ that carries the `Coulombic information' in $\re \Phi_{1}^{0}$ that escapes the radiative modes $A_{2}^{0}$.} Note, in particular, that the electric charge $Q$ can also be expressed as 
\be \label{charge2} Q \= - \f{1}{2\pi}\oint G(\theta,\varphi) \rmd^{2} S. \ee

\item From Peeling properties (\ref{peeling2}) and (\ref{peeling3}) it follows that the field $A_{1}^{0} - \phi^{0}$ is constant on $\scrip$ and its value does not enter any of the physical quantities one normally considers, such as energy, momentum, angular momentum, electric charge, infra-red charges \cite{aa-bib} and electromagnetic memory effect \cite{bg}. These conclusions can also be reached directly from Maxwell equations. Indeed, in section \ref{s3}, where we discuss the gravitational case, we will complement the approach to Peeling we used here by obtaining the analogous relations directly from linearized Einstein's equations. But as we will see, calculations become cumbersome. 

\item In the Coulomb gauge the vector potential satisfies $\Do^{a} A_{a} =0$ everywhere in Minkowski space. Multiplying this equation by $\Omega^{-1}$ and taking the limit to $\scrip$, we obtain:
\be \partial_{u} A_{1}^{0} \=0 \, . \label{coulomb1}\ee
Since $(\phi^{0} - A_{1}^{0})$ is constant on $\scrip$, we conclude that $\partial_{u} \phi^{0} \= 0$ as well. Maxwell's equation (\ref{max1}) now implies that the total electric charge of the system is directly determined by the scalar potential $\phi^{0}$ on $\scrip$:
\be \label{charge3} Q = \f{1}{4\pi} \oint \phi^{0}(\theta,\varphi) \rmd^{2} S \ee
where the integral is taken over a cross-section of $\scrip$.

\item Finally, we observe that several simplifications occur \emph{in absence of  sources}. Let us work in the Coulomb gauge. Then, Eq.~(\ref{max1}) leads us to set $\phi=0$ everywhere in Minkowski space. In particular this implies $\phi^{0} = 0$,\, and $\phi^{1} =0$ in Eq.~(\ref{Phi}) (and that we can set $A_{a}\h{n}^{a} \= 0$). Furthermore, now the Coulomb gauge and Lorenz gauge coincide. Next, multiplying the Transversality condition $\Do^{a} A_{a}=0$ by $\Omega^{-1}$ and $\Omega^{-2}$ and taking the limit to $\scrip$, we obtain, respectively, (\ref{coulomb1}) and 
\be \label{coulomb2} \partial_{u} A_{1}^{1} \= 2\sqrt{2} \re (\eth A_{2}^{0}) + A_{1}^{0}\, .\ee
Furthermore, in the source-free case, now under consideration, the limit to $i^{o}$  along $\scrip$ of $A_{1}^{0}$ vanishes if the initial data for $\vec{A}_{a}$ falls off at spatial infinity sufficiently rapidly. Since we already know that $A_{1}^{0}$ is a constant on $\scrip$, it is identically zero. Using vanishing of $\phi^{0}, \, \phi^{1}$ and $A_{1}^{0}$ and (\ref{coulomb2}) in the second equation in (\ref{Phi}) we obtain:
\be \label{phi1}
\Phi_1^{0} = \frac{\sqrt{2}}{r^{2}}\, \eth A_{2}^{0}  \qquad {\hbox{\rm whence \quad $ G(\theta,\varphi) \= 0$  in (\ref{rephi10})\, . }}\ee 
We will find that this simplification makes a key difference in the consideration of angular momentum with and without source currents.\\% Complications introduced by the presence of sources were highlighted by Racz in \cite{racz}.\\ 
\end{itemize}

Let us summarize. In absence of sources, there are natural gauges in which the 4-potential $A_{a}$ admits a smooth limit to $\scrip$. Furthermore, one can choose a gauge with $A_{a}\hat{n}^{a}\=0$. Then the two remaining components of the pull-back $\pb{A_{a}}$ of $A_{a}$ to $\scrip$ are encoded in $A_{2}^{0}$ \cite{aa-bib}. These radiative modes encode full information in the source-free solution. In particular, the energy, momentum and angular momentum carried by electromagnetic fields can be expressed using $A_{2}^{0}$\, \cite{aams}. The situation is quite different once we have sources. The 4-potential can diverge at $\scrip$ in well-motivated gauges (such as the Coulomb and the Lorenz gauge). Now, the components $A_{2}^{0}$ at $\scrip$ carry information only about the two radiative degrees of freedom in the solution. The `Coulombic information' is encoded in other components of the potential, e.g. through $G(\theta,\varphi)$.

\subsection{$A_{a}^{\T}$ versus $A_{a}^{\t}$}
\label{s2.3}

Sections \ref{s2.1} and \ref{s2.2} provide the necessary platform to  compare the two notions of transversality. We will first introduce these notions, emphasizing their asymptotic behavior, and then contrast them.

\subsubsection{The two notions}
\label{s2.3.1}

By definition $A_{a}^{\T}$ satisfies the Coulomb (or Transverse) gauge condition $\Do^{a}\vec{A}_{a}^{\,\T}=0$ everywhere in Minkowski space. Since by assumption $A_{a}^{\T}$ must satisfy (\ref{falloff1}) - (\ref{falloff3}), it follows that there is no residual gauge freedom. Hence all expressions constructed from $\vec{A}_{a}^{\,\T}$ are gauge invariant on the entire space-time.

\emph{For notational clarity, from now on we will use an underbar to denote vector potentials in the Coulomb gauge.} The asymptotic expansions (\ref{falloff1}) - (\ref{falloff3}) show that the leading-order part of $\Ac_{a}$ is captured in 4 functions on $\scrip$, namely, two real functions $\phic^{0}, \Ac_{1}^{0}$ and a complex function $\Ac_{2}^{0}$. As we saw, $\ub{A}_{2}^{0}$ represents the two radiative degrees of freedom of the Maxwell field. At first, it seems surprising that the leading-order part of the 3-potential, $\vec{\Ac}_{a}$, has an additional component $\Ac_{1}^{0}$. Shouldn't the requirement of Transversality leave us with just two? Recall from (\ref{coulomb1}) that the Transversality condition $\Do^{a}\vec\Ac_{a}=0$ does lead to a non-trivial restriction: $\Ac_{1}^{0}$ is non-dynamical; it is a function only of $\theta$ and $\varphi$ on $\scrip$. That is, $\vec{\Ac}_a$ carries information worth two (real) functions $\Ac_{2}^{0}(u,\theta,\varphi)$ of three variables \emph{and} one function $\Ac_{1}^{0}(\theta,\varphi)$ only of two variables. \emph{Thus, in terms of asymptotic fields on $\scrip$, implications of Transversality are subtle:} Rather than leaving us with only two radiative modes $\Ac_{2}^{0}$, it also provides an additional and unanticipated non-dynamical function $\Ac_{1}^{0}$ of two variables. Now, as we saw in section \ref{s2.2}, because of Maxwell's equations (which in particular imply the Peeling properties),\, $(\phic^{0} - \Ac_{1}^{0})$ is a constant on $\scrip$ in any gauge. Hence in the Coulomb gauge now under consideration, $\phic^{0}$ is also a non-dynamical function only of $\theta,\varphi$ and we can trade $\Ac_{1}^{0}$ for $\phic^{0}$. We will do so because, in view of (\ref{max1}), the interpretation of $\phic^{0}$ is transparent: it directly captures the `Coulombic aspects' of the asymptotic Maxwell field. In particular, as we saw in Eq (\ref{charge3}), its integral over a 2-sphere cross-section of $\scrip$ provides the electric charge. Finally, since the 4-potential $\ub{A}_{a}$ is gauge invariant on the entire space-time, higher order fields such as $\phic^{1}$ and $\Ac_{1}^{1}$ also have an invariant meaning. They feature in the expression (\ref{Phi}) of $\re \Phi_{1}^{0}$ and, together with $\phic^{0}$ and $\Ac_{2}^{0}$, suffice to determine it. Therefore, the function $G(\theta,\varphi)$ in (\ref{rephi10}) is also determined by these fields. This fact will play an important role in our discussion of angular momentum. \\ 

The second notion of transversality $A_{a}^{\t}$ that is widely used in the literature (to motivate constructions in the gravitational case) is local in physical space (see, e.g. \cite{pw}). One sets
\be \label{At}
A_a^{\t} := P_a^{\, b} A_b, \quad {\rm where}\quad P_{a}^{\,b} = m_a \bar{m}^b + \bar{m}_a m^b\,\, \equiv\,\, \hat{m}_a \hat{\bar{m}}^b + \hat{\bar{m}}_a \hat{m}^b \ee
is the projection operator that projects fields into the 2-spheres $r={\rm const},\, t = {\rm const}$ in Minkowski space-time. Using the expansion (\ref{eq:vecexp}) of $A_{a}$ in terms of its components and the assumed  fall-off (\ref{falloff1}) - (\ref{falloff3}) of these components, we obtain the following expansion of $A_{a}^{\t}$ in a neighborhood of $\scrip$:
\be
A_a^{\t} = \Big(A_2^0 + \Omega A_2^1 +  \ldots \Big) \hat{m}_a
+ \Big(\bar{A}_2^0 + \Omega \bar{A}_2^1  + \ldots \Big) \hat{\bar{m}}_a 
\,\, \=  \,\, A_2^0 \, \hat{m}_a + \bar{A}_2^0 \, \hat{\bar{m}}_a \, ,
\ee
where, as before, $\=$ denotes equality \emph{at} $\scrip$. Note that $A_{a}^{\t}$ automatically satisfies $A_{a}^{\t}\h{n}^{a} \=0$ at $\scrip$.

Sometimes $A_{a}^{\t}$ is defined via (\ref{At}) without specifying any gauge conditions even though it is obvious that the result is not gauge invariant. What would happen if we use the Lorenz gauge in Minkowski space-time so that the dynamical equation satisfied by $A_{a}$ is just the wave equation? Indeed, more careful treatments make this choice. Then the residual gauge freedom is restricted to $A_a \to A_a + \grad_a \Lambda$ with $\Box \Lambda = 0$. Since this gauge transformation also needs to preserve the fall-off conditions (\ref{falloff1}) - (\ref{falloff3}), the solution to the wave equation has the form%
\footnote{Note that if $\Lambda$ had a leading order term of the type $\Lambda = \Lambda^{0}(\theta,\varphi) + \ord(\f{1}{r})$, the Cartesian components of $\nabla_{a}\Lambda$ \emph{would} fall-off as $1/r$ as needed.  But this possibility is ruled out by the fact that $\Box \Lambda$ would then not vanish to leading order.
If $\Lambda$ had terms of the form $\Lambda= {(\ln r/r) \, {\Lambda}^{0}(u,\theta,\varphi)} + \ord(\f{1}{r^{2}})$, or $\Lambda= {\Lambda}^{0}(u,\theta,\varphi) + \ord(\f{1}{r})$, the gauge transformation would fail to preserve the fall-off conditions on $A_{a}$.}
\be
\Lambda = \frac{\Lambda^{0}(u,\theta,\varphi)}{r} + \ord\Big(\f{1}{r^{2}}\Big)\, ,
\ee
leading to gauge transformation on $A_a$ of the form
\be 
A_a  \; \to \; A^{\prime}_a= A_a + \frac{\del_u \Lambda^{0}(u,\theta,\varphi) }{r} \grad_a u + \frac{\del_\theta \Lambda^{0}(u,\theta,\varphi) }{r} \grad_a \theta + \frac{\del_\varphi \Lambda^{0}(u,\theta,\varphi) }{r} \grad_a \varphi + \ord\Big(\f{1}{r^{2}}\Big) \, .\nonumber
\ee
Hence, 
\be \phi^{0} \to \phi^{0} - {\del_u \Lambda^{0}(u,\theta,\varphi) },\qquad A_{1}^{0} \to A_{1}^{0} - {\del_u \Lambda^{0}(u,\theta,\varphi) }, \qquad A_{2}^{0} \to A_{2}^{0}\, .\ee 
Since $A_{a}^{\t} = A_{2}m_{a} + A_{2}\bar{m}_{a}$, we have: $A_{a}^{\t} \hat{=} {A}_a^{\prime\, \t}$\, at\, $\scrip$. Thus, $A_a^{\t}$ is gauge invariant on $\scrip$ if $A_a$ satisfies the Lorenz gauge near $\scrip$ and has the assumed fall-off behavior. Note however that, even with the Lorenz gauge imposed, $\phi^{0}$ and higher order fields (including $\phi^{1}$ and $A_{1}^{1}$) that enter the expression of $\Phi^{0}_{1}$ in Eq. (\ref{Phi}) are not gauge invariant. \emph{For notational clarity, from now on we will denote fields associated with the vector potential in the Lorenz gauge by an undertilde.} Thus, $\ut{A}_{2}^{0}$ will denote the radiative modes extracted from $A_{a}^{\t}$ in the Lorenz gauge.

\subsubsection{Comparison}
\label{s2.3.2}

We can now compare the two notions of transversality. There are three important differences. First, even with the Lorenz gauge condition, $A_{a}^{\t}$ is \emph{not} gauge invariant beyond the leading asymptotic order. By contrast, $A_{a}^{\T}$ is fully gauge invariant. The second difference has to do with the leading order fields. As we saw, each notion of transversality enables one to single out two radiative modes at $\scrip$: $\Ac_{2}^{0}$ from $A_{a}^{\T}$, and $\ut{A}_{2}^{0}$ from $A_{a}^{\t}$. Since the Transversality condition on $\vec{\Ac}_{a}$ fixes the gauge completely, there is no gauge freedom in the modes $\Ac_{2}^{0}$. Similarly, $\ut{A}_{2}^{0}$ is invariant under the restricted gauge transformations compatible with the Lorenz gauge. Furthermore, we know that the gauge invariant field $\Phi_{2}^{0}$ is related to the vector potential via $\Phi_{2}^{0} = \sqrt{2}\,\partial_{u}\, A_{2}^{0}$  \emph{in any gauge} satisfying our fall-off conditions. Therefore, the radiative modes in the two notions are related by a non-dynamical function $f(\theta,\varphi)$ on $\scrip$:
\be \Ac_{2}^{0}(u,\theta,\varphi) - \ut{A}_{2}^{0}(u,\theta,\varphi) \,\= \, f(\theta,\varphi)\, .
\ee
(Since $\im \Phi_{1}^{0} \= \sqrt{2}\,\, \im (\eth A_{2}^{0})$ in any gauge, it follows that $\im (\eth f) \=0$  which in turn implies that $f= \bar\eth h$ where $h$ is real, i.e., $f$ is `purely electric' \cite{np}.) There is no a priori guarantee that the functions representing radiative modes in $A^{\T}_{a}$ are the same as those in $A_{a}^{\t}$ \emph{even at $\scrip$.} However, the difference between them is a non-dynamical function. As we show below, it drops out of all physical quantities that \emph{can be constructed from the radiative modes} --including those associated with soft (i.e. infrared) `charges' and the memory effect. 
 
Third, the main difference in the two approaches to transversality of the vector potential is the following. In the first approach, $\vec{\Ac}_{a} = \vA_{a}^{\,\,\T}$ has \emph{three} independent components at $\scrip$ even to leading order: In addition to $\Ac_{2}^{0}$, the leading-order part of $\vec{\Ac}_{a}$ also provides us with $\Ac_{1}^{0}$, which can be traded for $\phic^{0}$. This additional field is non-dynamical on $\scrip$, a hallmark of fields carrying the `Coulombic information' in the asymptotic Maxwell field. Furthermore, in the $A_{a}^{\T}$ framework higher order fields such as $\phic^{1}$ and $\Ac_{1}^{1}$ are also accessible and gauge invariant. As Eq. (\ref{Phi}) shows, we can use them to construct the component $\re\,\Phi_{1}^{0}$ of the asymptotic Maxwell field that carries `Coulombic information'. By contrast, in the second approach,  $A^{\t}_{a}$ provides us with only two components $\ut{A}_{2}$ of the vector potential and they do not suffice to determine $\Phi_{1}^{0}$.

How does this difference manifest itself? First, as we saw, the electric charge cannot be recovered from the two radiative modes. However, it can be expressed as an integral of $\re\,\Phi_{1}^{0}$ or
$\phic^{0}$ over a 2-sphere cross-section of $\scrip$, directly accessible in the $A_{a}^{\T}$ approach. This is not possible in the $A_{a}^{\t}$ approach. Thus, the information about leading order `Coulombic properties' of the solution is accessible to the first approach but not to the second. Let us consider two simple examples that bring out this point:
\begin{itemize}
\item For a charge $\mathfrak{q}$ moving in the $x-$direction with velocity $v$ in the frame in which the Coulomb gauge is imposed, we have
\be \label{pointcharge}
\phic^{0} =  \frac{\mathfrak{q}}{\sqrt{1 - 2 v \sin \theta \cos \phi + v^2}}, \qquad
\Ac_1^0 = -\mathfrak{q} +  \frac{\mathfrak{q}}{\sqrt{1 - 2 v \sin \theta \cos \phi + v^2}},\qquad \Ac_{2}^{0} =0\, .
\ee
\item For an oscillating dipole situated at the origin with strength $\mathfrak{p}$ and frequency%
\footnote{\label{osc}In this example, we envisage that the dipole has these periodic oscillations only for a finite duration and focus just on the corresponding interval of retarded time near $\scrip$.}
$\omega$ so that the source is given by $j_a = \frac{\mathfrak{p}}{4\pi} \left(\cos \omega t \, \delta(x) \delta(y) \delta'(z) \grad_a t - \omega \sin \omega t \, \delta^{(3)}(\vec{x}) \grad_a z \right)$, in the Coulomb gauge one obtains \cite{ep}
\be \label{dipole} \phic^{0} =  0  \qquad  \Ac_1^0 = 0 \qquad \Ac_2^0 = \bar{\Ac}_2^0 = \frac{\mathfrak{p} \, \omega}{4\sqrt{2} \pi} \sin \theta \, \sin \omega u\, .
\ee
\end{itemize}
In the first case $A^{T}_{a}$ carries additional `Coulombic information' that is absent in $A^{t}_{a}$ while in the second case the leading order potentials do not carry any `Coulombic information' because the total charge vanishes. In the gravitational case, the situation is parallel, with the electric charge replaced by the linearized Bondi 4-momentum.

Given this difference, one's first intuition could be that the $A_{a}^{\t}$-approach would not be adequate to handle issues such as the soft charges and memory effect which are related to the `charge aspect' $\Phi_{1}^{0}$ at $\scrip$, but should be adequate for calculating physical quantities associated with electromagnetic waves such as the energy, momentum and angular momentum they carry. \emph{However, it turns out that neither of these expectations is borne out; the situation is more subtle.}

\subsubsection{Fluxes of energy-momentum and angular momentum}
\label{s2.3.3}

Let us begin with the fluxes of energy-momentum and angular momentum across $\scrip$. Since every Killing field $K^{a}$ admits a smooth limit (\ref{killing}) to $\scrip$, using conformal invariance of the Maxwell field, the flux $\mathcal{F}_{K}$ can be expressed as:
\ba \label{kflux}
\mathcal{F}_{K} &=& \int_{\scrip} \hat{T}_{ab}\, K^{a} \ti{n}^{b}\, {\rmd} u\, {\rmd^2}{S}\nonumber\\ 
&=& \int_{\scrip} \big(\h{F}_{ac}\h{F}_{bd}\,\h\eta^{cd}\, - \f{1}{4} \h\eta_{ab} \h{F}_{cd} \h{F}^{cd}\,\big) K^{a}\ti{n}^{b}\, {\rmd} u\, \rmd^2{S}\, . \ea
Recall from (\ref{killing}) that for translation Killing fields, $K^{a} \= \alpha(\theta,\varphi) \ti{n}^{a}$ (where $\alpha(\theta,\varphi)$ characterizing the translation $K^{a}$ is a linear combination of the first four spherical harmonics). Therefore, the corresponding flux becomes:
\ba \label{emflux} \mathcal{F}_{\alpha} &=& \int_{\scrip}  |\Phi_{2}^{0}|^{2} \, \alpha(\theta,\varphi)\, {\rmd} u\, {\rmd^2}{S}\nonumber\\
&=& \int_{\scrip} 2\, |\partial_{u} A_{2}^{0}|^{2}\, \alpha(\theta,\varphi)\,{\rmd} u\, {\rmd}^2{S} \, .\ea
Thus the energy-momentum flux is expressible entirely in terms of the radiative modes $A_{2}^{0}$. Although we focused on the total flux of energy-momentum across $\scrip$, since no integration by parts was involved, it is clear from the calculation that the integrand, representing the \emph{local flux}, can also be expressed in terms of $A_{2}^{0}$. Furthermore, the expression holds in any gauge.\\

Next, let us consider the component of angular momentum along a spatial rotation $h^{a}$ which is tangential to the $u={\rm const}$ 2-sphere cross sections of $\scrip$  and satisfies $\mathcal{L}_{h} q_{ab} \= 0$ and $\mathcal{L}_{h} \ti{n}^{a}  \=0$ (see (\ref{killing})). Then we can expand $h^{a}$ as $h^{a} \= g(\theta,\varphi) \h{m}^{a} +  \bar{g}(\theta,\varphi) \h{\bar{m}}^{a}$ (where $g$ satisfies $\bar\eth {g} \= 0$ \cite{tdms,ak-thesis}). Substituting $K^{a} = h^{a}$ in (\ref{kflux}), we obtain the flux of the $h^{a}$-component of angular momentum:
\ba \label{amflux1} \mathcal{F}_{g} &=& \sqrt{2} \int_{\scrip} \re\big[ 
\bar\Phi_{2}^{0}\,\,\Phi_{1}^{0} \, g(\theta,\varphi)\big]\, {\rmd} u\, {\rmd^2}{S}\\
&=& 2 \int_{\scrip} \re\big[ (\partial_{u} \bar{A}_{2}^{0})\, (\sqrt{2} \eth A_{2}^{0} +G(\theta,\varphi))\, g(\theta,\varphi)\big] \, {\rmd} u\, {\rmd^2}{S} \label{amflux2} 
% \\ &=& 2 \int_{\scrip} \re\big[ (\partial_{u} \bar{A}_{2}^{0})(\sqrt{2} \eth A_{2}^{0})\,g(\theta,\varphi)\big] \,- \, \Big(\oint_{u=\infty} - \oint_{u=-\infty}\Big)\, \re [\bar{A}_{2}^{0} \, G(\theta,\varphi)\, g(\theta,\varphi)] 
\ea
where in the second step we have used (\ref{rephi10}) to express the real part of $\Phi_{1}^{0}$ in terms of $A_{2}^{0}$ and $G(\theta,\varphi)$. Again, these expressions hold in any gauge. The integrand in (\ref{amflux1}) involves only $\Phi_{2}^{0}$ and $\Phi_{1}^{0}$ which are both manifestly gauge invariant%
\footnote{While the second expression also holds in any choice of gauge since $\partial_{u} A_{2}^{0}$ and $\re \Phi_{1}^{0} = \sqrt{2} \re\eth A_{2}^{0} + G(\theta,\varphi)$ are both gauge invariant, the individual terms, $\re \eth A_{2}^{0}$ and $G(\theta,\varphi)$ are not. In particular $\eth \Ac_{2}^{0}$ that features in the expression of $A_{a}^{\T}$ need not equal $\eth \ut{A}_{2}^{0}$ that features in $A_{a}^{\t}$.}
and the passage to (\ref{amflux2}) featuring $A_{2}^{0}$ and $G(\theta,\varphi)$ did not involve any integration by parts. Therefore the integrands in each of these expressions represents the \emph{local flux} of angular momentum. In contrast to energy-momentum, the local as well as the integrated flux of angular momentum  depends on the asymptotic `Coulombic part' of the Maxwell field through $\re \Phi_{1}^{0}$. Eq.~(\ref{amflux2}) makes it explicit that the flux of angular momentum \emph{cannot} be expressed purely in terms of the two radiative modes captured in $A_{2}^{0}$; we also need the `Coulombic information' in $\re\Phi_{1}^{0}$ encoded in the function $G(\theta,\varphi)$. 

In particular, then, angular momentum carried away by electromagnetic waves \emph{cannot be expressed using only the two components $\ut{A}_{2}^{0}$ of the electromagnetic vector potential ${A}^{\t}_{a}$.} This fact appears not to have been noticed in the literature presumably because this feature arises only in presence of sources; as we saw in Eq. (\ref{phi1}), in the source-free case $G(\theta,\varphi) \= 0$. By contrast, $A_{a}^{\T}$ provides additional asymptotic fields that are sufficient to obtain $\re\Phi_{1}^{0}$ (and $G(\theta, \varphi)$) and hence the local and total flux of angular momentum at $\scrip$ also in presence of sources.

Perhaps the simplest illustration of this limitation of the $A_{a}^{\t}$ framework is provided by a linear superposition of the Coulomb field of a static point charge (obtained by setting $v=0$ in (\ref{pointcharge})) and the oscillating dipole of (\ref{dipole}). For the point charge, the radiation field $\Phi_{2}^{0}$ and the radiative modes $A_{2}^{0}$ vanish identically. Therefore the local angular momentum flux also vanishes identically. However, $\re\Phi_{1}^{0}$ does not vanish as it carries the Coulombic information in the solution, whence $G(\theta,\varphi)$ is non-zero, given by $G(\theta, \varphi) = - \mathfrak{q}/2$. For the oscillating dipole, the situation is almost the opposite: Now $\Phi_{2}^{0},\, A_{2}^{0}$ and the local angular momentum flux are all non-zero, but $\re \Phi_{1}^{0}$ is completely determined by $A_{2}^{0}$, whence $G(\theta, \varphi) =0$. Therefore the flux of angular momentum can be expressed entirely in terms of $A_{2}^{0}$. However, for the superposed solutions none of the fields vanish. Since, in particular, $G(\theta, \varphi) \not=0$, the local flux of angular momentum at $\scrip$ is no longer expressible purely in terms of $A_{2}^{0}$. Thus, even though the pure Coulomb solution of a static charge does not carry any angular momentum, in the superposition it modifies the expression of local angular momentum flux in a non-trivial fashion. Consequently, while we can express this flux using the asymptotic fields $A_{a}^{\T}$, we cannot express it using $A_{a}^{t}$ alone.

This simple example brings out the essence of this phenomenon. Suppose we are given a general retarded solution $A_{a}$ such that $G(\theta,\varphi)=0$, i.e. $\re \Phi_{1}^{0} = \sqrt{2} \, \re \eth A_{2}^{0}$. Then angular momentum flux (\ref{amflux2}) at $\scrip$ can be expressed purely in terms of $A_{2}^{0}$. Note, however, that because $G(\theta, \varphi) =0$, the total charge of this system is zero (see Eq. (\ref{charge2})). Now let us superimpose on this solution a static Coulomb field so that the total charge is non-zero. Then the angular momentum flux (\ref{amflux2}) at $\scrip$ is no longer expressible purely in terms of $A_{2}^{0}$; it is not accessible in the $A_{a}^{\t}$ framework but continues to be accessible in the $A_{a}^{\T}$ framework. Thus, the subtlety is directly related to the `Coulombic information' in $\re \Phi_{1}^{0}$ in presence of sources carrying a non-zero charge.

\subsubsection{Soft charges and electromagnetic memory}
\label{s2.3.4}
 
The `soft' charges associated with the Maxwell field play an important role in the discussion of the $S$-matrix in quantum electrodynamics that adequately handles infrared issues in the photon sector (see, e.g., \cite{fk,aakn,krakow}). For our purposes, it will suffice to simply note that they are obtained by integrating the charge aspect $\re\Phi_{1}^{0}$ against real-valued test fields $\ti\alpha (\theta,\varphi)$ \cite{aa-bib}:
\be \label{soft1}  q_{\ti\alpha} = \f{1}{4\pi}\,\Big(\oint_{u=\infty} - \oint_{u=-\infty}\Big)\, \big[\ti\alpha (\theta,\varphi)\,\, \re\Phi_{1}^{0}(u,\theta,\varphi)\big]\, \rmd^{2}S.\ee
For any Maxwell field, by charge conservation $q_{\ti\alpha}$ vanishes if $\ti\alpha ={\rm const}$. However, for a general $\ti\alpha $ it carries non-trivial information about the `Coulombic aspect' of the asymptotic Maxwell field. Since the integrand features $\re\Phi_{1}^{0}$, one might first expect that the soft charge would not be expressible purely in terms of radiative modes. However, using (\ref{rephi10}), it follows that 
\be \label{soft2} q_{\ti\alpha} = -\f{\sqrt{2}}{4\pi}\, \Big(\oint_{u=\infty} - \oint_{u=-\infty}\Big)\, \re \big[A_{2}^{0}(u,\theta, \varphi)\,\, \eth \ti\alpha(\theta,\varphi)\big]\, \rmd^{2}S\, , \ee
\emph{since $G$ is $u$-independent.} Thus, the soft charges can in fact be expressed using only the radiative modes in $A_{2}^{0}$. (Since the kernel of $\eth$ on spin-weight zero functions consists only of constants, in general $q_{\ti\alpha}$ vanishes only if $\ti\alpha= {\rm const}$.) Indeed, using the Maxwell equation $\partial_{u}\Phi_{1}^{0} = \eth \Phi_{2}^{0}$ we can recast the soft charges in terms only of the radiation field $\Phi_{2}^{0}$:
\ba \label{soft2} q_{\ti\alpha} &=& -\f{1}{4\pi}\,\int_{\scrip} \big(\Phi_{2}^{0} (u,\theta,\varphi)\,\,\eth \ti\alpha\big)\, \rmd u\, \rmd^{2}S\nonumber \\
&=& - \f{1}{4\pi}\,\oint \rmd^{2}S \,\eth \ti\alpha\,\, \Big(\int_{-\infty}^{\infty} \rmd u\, \Phi_{2}^{0}(u,\theta,\varphi)\Big)\, .
\ea
Thus, even though the soft charges are associated with $\re\Phi_{1}^{0}$ that one normally thinks of as carrying `Coulombic information', they can be expressed entirely in terms of radiative modes and are therefore accessible in the $A^{t}_{a}$-framework. We will now show that the situation is similar for electromagnetic memory.\\
 
The electromagnetic analog of the memory effect is a `kick' --i.e., change in velocity $v$ that a test particle with charge $\mathfrak{q}$ in the asymptotic region undergoes after the passage of an electromagnetic wave. If the particle is initially following a trajectory of the time-translation Killing field $t^{a}$ in the asymptotic region with $\theta=\theta_{o}, \varphi= \varphi_{o}, r= r_{o}$, it would acquire a velocity of magnitude $\Delta v$ given by \cite{bg}%
\be \Delta v = \frac{\mathfrak{q}}{m} \left[{\int_{-\infty}^{\infty} \rmd u \vec{E}(u,r_{o},\theta_{o},\varphi_{o})  \, \cdot \, \int_{-\infty}^{\infty} \rmd u \vec{E}(u,r_{o},\theta_{o},\varphi_{o})}\right]^{\f{1}{2}} 
,\ee
where $E^{a}$ is the electric field $E^{a} = F^{ab}t_{b}$ in the rest frame of the particle. We can rewrite this expression using the  asymptotic expansion of the Maxwell field as:

\begin{align}
\Delta v &= \frac{\mathfrak{q}}{m r_{o}} \Big{|}\int_{-\infty}^{\infty}\rmd u\,\, \Phi_{2}^{0}(u,\theta_{0},\varphi_{0}) \Big{|} \,+\, \ord\Big(\f{1}{r_{o}^{2}}\Big)\nonumber\\ 
&= \frac{\sqrt{2} \mathfrak{q}}{m r_{o}} \left| A_2^0(u=\infty,\theta_{o},\varphi_{o}) - A_2^0(u=-\infty,\theta_{o},\varphi_{o}) \right| + \ord \Big(\f{1}{r_{o}^{2}}\Big)\, .
\end{align}
Thus, the leading term in the electromagnetic memory is proportional to the absolute value $|q_{\ti\alpha}|$ of the soft charge, for $\ti\alpha$ such that $\eth \ti\alpha$ is the Dirac distribution centered at $\theta =\theta_{o}$ and $\varphi=\varphi_{o}$. %Although $\Delta v$ is closely related to the `charge aspect' $\re\Phi_{1}^{0}$, 
It can can be expressed entirely in terms of the radiation field $\Phi_{2}^{0}$, or equivalently, the radiative modes $A_{2}^{0}$ of the vector potential. \\

\subsubsection{Summary}
\label{s2.3.5}

$A_{a}^{\T}$ is gauge invariant everywhere in space-time while only the limit to $\scrip$ of $A_{a}^{\t}$ is gauge invariant. At $\scrip$, the $A_{a}^{\T}$-framework provides us two leading order fields  $\Ac_{2}^{0}$ and $\phic^{0}$ (or, equivalently, $\Ac_{1}^{0}$), as well a hierarchy of higher order fields, such as $\phic^{1},\, \Ac_{1}^{1},\, \ldots$ that are all gauge invariant. By contrast, the only  gauge invariant field that the $A_{a}^{\t}$-framework provides at $\scrip$ is $\ut{A}_{2}^{0}$. The two radiative modes of the electromagnetic field are encoded in $\Ac_{2}^{0}$ and $\ut{A}_{2}^{0}$. However, in general $\Ac_{2}^{0}$ and $\ut{A}_{2}^{0}$ do not agree even asymptotically, i.e., at $\scrip$, but their difference is encoded in a non-dynamical function $f(\theta,\varphi)$ that is irrelevant for all physical observables. We can express energy-momentum, soft charges and electromagnetic memory using only the radiative modes $\Ac_{2}^{0}$ or $\ut{A}_{2}^{0}$. However, in presence of sources, this is not the case for angular momentum carried by electromagnetic waves. One's first intuition could be that, since physical observables that can be expressed entirely in terms of the field strength $F_{ab}$ cannot depend on the choice of gauge, it should be possible to express angular momentum flux $\mathcal{F}_{g}$ using either of the two methods. However, the expression of  $\mathcal{F}_{g}$ involves $\re\Phi_{1}^{0}$
which, although gauge invariant, cannot be expressed in terms of $A_{a}^{\t}$. By contrast, additional fields at $\scrip$ provided by the $A_{a}^{\T}$-framework do carry this information.\\
\goodbreak

\emph{Remark:}\\
It is instructive to rewrite the angular momentum flux  (\ref{amflux2}) by performing an integration by parts as:
\be \mathcal{F}_{g} = 2 \int_{\scrip} \re\big[ (\partial_{u} \bar{A}_{2}^{0})(\sqrt{2} \eth A_{2}^{0})\,g(\theta,\varphi)\big] \,- 2\, \Big(\oint_{u=\infty} - \oint_{u=-\infty}\Big)\, \re [\bar{A}_{2}^{0} \, G(\theta,\varphi)\, g(\theta,\varphi)]\, . \label{amflux3} \ee
This form brings out the fact that while the flux does depend on the `Coulombic information' in $\re \Phi_{1}^{0}$, it really enters only through boundary terms at the two ends, $i^{+}$ and $i^{o}$, of $\scrip$. We emphasize again that the subtlety associated with the angular momentum flux arises only in presence of sources, since we know from (\ref{phi1}) that $G(\theta,\varphi)$ vanishes in the source-free case.

\section{Linearized Gravity}
\label{s3}

The overall situation for linearized gravitational fields parallels that in the Maxwell case. But the mere fact that we now have a second rank tensor field $h_{ab}$ in place of the 1-form $A_{a}$ makes calculations significantly longer and less transparent. However, we will be able to use the fact that the underlying conceptual structure is very similar to that in the Maxwell case to shorten and streamline the discussion and see our way through structures that appear opaque and complicated at first. 

This section is divided into three parts. In the first we will use the Transverse gauge that is well adapted to the $\TT$-notion to arrive at  the `minimal' fall-off conditions on metric perturbations $h_{ab}$ as one recedes from sources in null directions. Using these results as motivation, we will specify the class of metric perturbations (i.e., gauges) we wish to consider. In the second part we discuss the consequences of linearized Einstein's equations on these metric perturbations, including the Peeling properties of the Weyl tensor. Finally, by expressing these curvature scalars in terms of metric perturbations we isolate the radiative modes. In the third part we introduce the two notions $h_{ab}^{\TT}$ and $h_{ab}^{\tt}$ and compare and contrast them.

\subsection{Asymptotic conditions on metric perturbations $h_{ab}$}
\label{s3.1}
Since the notion of $h_{ab}^{\TT}$ is tied to the Transverse gauge, as in the Maxwell case we will use this gauge to motivate our asymptotic conditions on metric perturbations.

Consider a foliation of Minkowski space-time by $t={\rm const}$ hyperplanes and decompose the space-time metric perturbation $h_{ab}$ using this foliation:
\be \label{decomp} h_{ab} = 2\phi \nabla_{a}t \nabla_{b}t +2 \vA_{(a}\nabla_{b)}t + \vh_{ab}\ee
where, as in the Maxwell case, the arrows over $\vA_{a}$ and $\vh_{ab}$ emphasize that these fields are tangential to the $t={\rm const}$ surfaces, and the factor of $2$ has been introduced in front of $\phi$ to ensure agreement with the existing literature \cite{pw}. Denote as before the intrinsic positive definite metric on these surfaces by $\qo_{ab}$ and its torsion-free derivative operator by $\Do$. As in the Maxwell analysis, we start by assuming that $h_{ab}$ is smooth, its Cartesian components vanish at spatial infinity, and it is Transverse, i.e., satisfies $\Do^{a}\vh_{ab} =0$. Then the  restricted gauge freedom is given by
\be \label{restricted1} h_{ab} \to h_{ab} + \mathcal{L}_{\xi} \eta_{ab}, \qquad {\rm with} \qquad \xi^{a} = f t^{a} \ee
for some smooth space-time function $f$, where $t^{a}$ is the unit normal to the foliation. $\vh_{ab}$ is of course invariant under (\ref{restricted1}) while $\phi$ and $\vA_{a}$ transforms via
\be \phi \to \phi - \mathcal{L}_{t}f \equiv \phi - \dot{f} \quad {\rm and} \quad \vA_a \to  \vA_a - \Do_{a} f\, .  \ee
Note that the Transverse part $\vA_{a}^{\,\,\T}$ of $A_{a}$ is invariant under this restricted gauge freedom, and if we denote the longitudinal part by $\vA^{\,\L}_{a} \equiv \Do_{a} A^{\L}$, then $\Do_{a}(\phi - \dot{A}^{\L})$ is also gauge invariant. One can fix the restricted gauge freedom (\ref{restricted1}) in a number of ways, each yielding an equivalent but slightly different set of field equations. We will use this freedom to set $\vA_{a}^{\,\,\L} =0$ so that both $\vh_{ab}$ and $\vA_{a}$ are Transverse: $\Do^{a} \vh_{ab} =0$ and $\Do^{a} \vA_{a} =0$. 
These conditions exhaust the gauge freedom.
%There is still a small residual gauge freedom $h_{ab} \to h_{ab} + 2\nabla_{(a} f(t)\, t_{b)}$ \, where $f$ is now a function only of time. We will fix it by demanding that $\phi$ goes to zero as $r \to \infty$.
\\

\emph{Remarks:} 

(i) We can also use the restricted gauge freedom to set the 4-trace of $h_{ab}$ to zero: $h := \eta^{ab}h_{ab} = 0$\,\, or equivalently \,\, $2\phi = \qo^{ab} \vec{h}_{ab} \equiv \hb$. Note however that the restricted gauge freedom \emph{cannot} be used to set the 3-trace $\hb$ of our Transverse $\vec{h}_{ab}$ to zero because $\vec{h}_{ab}$ is invariant under (\ref{restricted1}). 

(ii) Note that our Transversality condition $\Do^{a} \vh_{ab} =0$ is distinct from what is sometimes called `Coulomb gauge' in the literature \cite{pw} in which $\vA_{a}^{\,\,\L}$ and certain parts of $\vh_{ab}$ (namely $S$ and $V^{\T}_{a}$ in Eq. (\ref{decomposition})) are set to zero.  In the `Coulomb gauge', $\Do^{a} \vh_{ab} \not=0$, but instead we have $\Do^{a} (\vh_{ab} - \f{1}{3} \bh \qo_{ab}) =0$. Therefore, we will refer to our choice only as \emph{Transverse gauge,} although in the Maxwell case we used the terms Transverse and Coulomb gauge interchangeably. In this section we work with the Transverse rather than Coulomb gauge both because the discussion is then completely parallel to that in the Maxwell theory, and because this gauge is generally not discussed in the literature. But the purpose of this section is only to motivate the fall-off conditions we want to use and the main results of this paper are the same in the two gauges.\\

To cast the linearized Einstein's equations in a 3+1 form, let us first perform a space-time decomposition of the stress-energy tensor:
\be \label{source}
\rho := t^a t^b T_{ab}, \quad \vec{J}_a :=  \qo_a^{\, b} \, t^c  \, T_{bc}, \quad  \vec{T}_{ab} := \qo_a^{\, m} \qo_b^{\,n} T_{mn}\, ,
\ee
and, as in the Maxwell case, further decompose the current $\vec{J}_a$ into its Transverse and Longitudinal parts:
\be \vec{J}_{a} = \vec{J}_{a}^{\,\T} + \vec{J}_{a}^{\, \L}, \quad {\rm where} \quad \Do^{a} \vec{J}_{a}^{\,\T} =0\, . \ee
As shown in Appendix \ref{a1}, we can now write the linearized Einstein's equations with (linearized) source as differential equations on the non-vanishing components of $h_{ab}$: $\phi,\, \vA\equiv \vA_{a}^{\, \T}$ and $\vh_{ab} \equiv \vh^{\, \T}_{ab}$. We obtain:
\ba
 \Do^{2} \phi &=& -4\pi G \,\,(\rho + \bar{T} - 2 \dot{J}^{\L})\label{phieq}\\
\Do^{2} \vA_{a}  &\equiv&  \Do^{2}\vA^{\,\,\T}_{a} = - 16\pi G \vec{J}_{a}^{\,\,\T} \label{Aeq}\\
 \Box \vh_{ab} &\equiv& \Box \vh^{\,\,\T}_{ab} = - 8\pi G \big(2\vec{T}_{ab} + (\rho- \bar{T})\, \qo_{ab}\big)^{{\T}}\, , \ea  
where $\bar{T} = \qo^{ab} \vec{T}_{ab}$,\,\, $\Do_{a} J^{\L} = \vec{J}_{a}^{\, \L}$\,\, and, as before the `dot' denotes the time derivative.
Thus, the `linearized lapse and shift fields' $\phi$ and $\vA_{a}$ are subject to elliptic equations while the spatial metric perturbation $\vh_{ab}$ is subject to a hyperbolic equation. Hence, the dynamical or propagating degrees of freedom are encoded in $\vh_{ab}$, just as one would expect. This is completely analogous to the situation in the Maxwell case. Again, we seek solutions to the Poisson equations which are well-behaved throughout space-time and go to zero at spatial infinity, and retarded solutions to the wave equation. 

As in the Maxwell case these equations, together with our assumptions on $T_{ab}$, lead us to conclude that it is appropriate to require that the Cartesian components of $h_{ab}$ fall-off as $1/r$. Therefore, when expanded in the basis vectors $\nabla_{a}t,\, \nabla_{a}r,\, m_{a}$ and $\bar{m}_{a}$,
\begin{align}
h_{ab} &= 2 \phi \, \grad_a t \, \grad_b t \, +\, 2 A_1 \, \grad_{(a} t \, \grad_{b)} r + 2 A_2 \, \grad_{(a} t \,  m_{b)} + 2 \bar{A}_2 \, \grad_{(a}t \, \bar{m}_{b)} \label{expn} \\
&\quad + B_{11} \, \grad_a r \, \grad_b r + 2 B_{12} \, \grad_{(a} r \, m_{b)} + 2 \bar{B}_{12} \, \grad_{(a} r \, \bar{m}_{b)} + B_{22} \, m_a \, m_b + \bar{B}_{22} \, \bar{m}_a \, \bar{m}_b + 2 C_{22} \, m_{(a} \, \bar{m}_{b)}\, , \notag
\end{align}
we will ask that the coefficients admit an expansion of the form
\begin{align}
&\phi = \frac{\phi^{0} (u, \theta,\varphi)}{r} + \frac{\phi^{1} (u, \theta,\varphi)}{r^2} + \frac{\phi^{(2)} (u, \theta,\varphi)}{r^3} +\ldots \nonumber\\
&A_1 = \frac{A_1^0 (u, \theta,\varphi)}{r} + \frac{A_1^1 (u, \theta,\varphi)}{r^2} + \frac{A_1^{(2)} (u, \theta,\varphi)}{r^3} + \ldots \label{falloff4} \\
& \text{similarly for $A_{2}, B_{11}, B_{12}, B_{22}$ and  $C_{22}$} \nonumber
\end{align}
Here $A_2$, $B_{12}$ and $B_{22}$ are complex-valued smooth functions (of appropriate spin weights) on $r={\rm const}$ surfaces and all other coefficients are real-valued and  smooth. To summarize, as in the Maxwell case, assuming that\\ (i) the physical stress-energy tensor $T_{ab}$ is smooth, has spatially compact support and its Cartesian components remain uniformly bounded in time, and,\\
(ii) the Cartesian components of $h_{ab}$ vanish at spatial infinity,\\
in the Transverse gauge we are led to the fall-off given in Eq. (\ref{falloff4}). As in the Maxwell case, we can rewrite the asymptotic conditions in a chart $u,\Omega, \theta,\varphi$ which is well-behaved at $\scrip$ (see Eq. \ref{asymexpn}). In the Maxwell case we found that in the Transverse gauge $A_{a}$ does not admit a well-defined limit to $\scrip$ already in simple examples, even though $F_{ab}$ is regular and smooth there. In the gravitational case, the situation is similar: With the fall-off (\ref{falloff4}), $\Omega^{2}h_{ab}$ does not have a well-defined limit at $\scrip$. Furthermore, this is the case even for the linearized gravitational field produced by a static point mass if one uses the Transverse gauge.\\ 

As in the Maxwell case, we use the Transverse gauge only as motivation to identify an appropriate class of gauges. Thus, \emph{we will now drop the restriction to the Transverse gauge and allow all gauges in which the coefficients $\phi, A_{1}, A_{2}, B_{11}, B_{12}, B_{22}$ and $C_{22}$ of the linearized metric $h_{ab}$ in the expansion (\ref{expn}) have the fall-off given by (\ref{falloff4}).} Metric perturbations $h_{ab}$ in this class have a rather weak fall-off. Of course there are gauges in which $\Omega^{2}\,h_{ab}$ is smooth at $\scrip$ (see, e.g., \cite{gx}). As in the Maxwell case, the main conceptual point is that in gauges we need to compare $h_{ab}^{\TT}$ and $h_{ab}^{\tt}$, the obvious strategy of demanding that $h_{ab}$ fall off such that $\Omega^{2} h_{ab}$ has a well-defined limit at $\scrip$ is not viable.

\subsection{Implications of field equations}
\label{s3.2}

In the Maxwell case, the Peeling properties ensured by the field equations on $F_{ab}$ provided additional constraints on leading order asymptotic fields. The situation is very similar in linearized gravity. Now the Peeling properties are implied by field equations satisfied by the linearized Weyl tensor \cite{rpwr}. One can therefore repeat the procedure we followed in the Maxwell case. 

However, there is the complementary approach that starts with field equations on $h_{ab}$ and arrives, among other consequences, at the Peeling properties. Each of these approaches brings out different aspects of the underlying structure. Since we used the first approach in the Maxwell case, for balance we will use the second approach for linearized gravity. We will not fix any gauge but only assume the asymptotic behavior specified in (\ref{falloff4}). Therefore, considerations of this section will apply to both notions, TT and tt, used in the literature.

\subsubsection{Conditions on leading order fields}
\label{s3.2.1}

Let us denote the linearized Einstein tensor by $G^{\prime}_{ab}$. It can be expressed in terms of the first order metric perturbation $h_{ab}$ as follows:
\be 2G^{\prime}_{ab} \equiv  - \Box h_{ab} + 2 \nabla_{(a}\nabla^{c} h_{b)c} - \nabla_{a}\nabla_{b} h + \big(\Box h - \nabla^{c}\nabla^{d} h_{cd}\big) \eta_{ab}\, , \ee %= 16\pi G T_{ab}\, , \ee
where $\nabla$ is the space-time derivative operator defined by $\eta_{ab}$. In the asymptotic region near $\scrip$ we will assume that source-free equations hold: $G^{\prime}_{ab}=0$.% 
\footnote{Actually, our discussion only assumes $G^{\prime}_{ab} = 8\pi G T_{ab}\,=\, O(1/r^{4})$.}
One can simply substitute the asymptotic expansion for $h_{ab}$ in the vacuum field equations and obtain conditions on coefficients at the leading and next to leading order in $1/r$:
\begin{align}
u^a u^b G_{ab}^{\prime} = 0 & \quad \Longrightarrow \quad \ddot{C}_{22}^0 = 0 \label{eq:uuprojection} \\
					& \quad \Longrightarrow \quad \dot{B}_{11}^0 - \dot{C}_{22}^0 + \ddot{C}_{22}^1 + 2 \sqrt{2} \re \eth \dot{B}_{12}^0 = 0\\
u^a r^b G_{ab}^{\prime} = 0 		& \quad \Longrightarrow \quad \dot{C}_{22}^0 - \sqrt{2} \re \left(\eth \dot{A}_2^0 + \eth \dot{B}_{12}^0 \right) = 0 \label{eq:urprojection}\\
u^a \bar{m}^b G_{ab}^{\prime} = 0 	& \quad \Longrightarrow \quad \ddot{A}_{2}^0 + \ddot{B}_{12}^0 = 0 \label{eq:umbarprojection1} \\
							& \quad \Longrightarrow \quad 
\dot{B}_{12}^0 + \f{1}{\sqrt{2}}\eth \dot{B}_{22}^{0} - \frac{1}{\sqrt{2}} \bar{\eth} \left(\dot{A}_1^0 + \dot{B}_{11}^0 + \dot{C}_{22}^0 \right) - \frac{1}{2} \left(\ddot{A}_2^1 + \ddot{B}_{12}^1 \right)= 0 \label{eq:umbarprojection2}\\
r^a r^b G_{ab}^{\prime} = 0 & \quad \Longrightarrow \quad 2 \dot{\phi}^0 + 2\dot{A}_{1}^0  + \dot{B}_{11}^0 = 0  \label{eq:rrprojection}\\
r^a \bar{m}^b G_{ab}^{\prime} = 0 & \quad \Longrightarrow \quad \dot{A}_{2}^0 +  \dot{B}_{12}^0 - \sqrt{2} \left(\bar{\eth} \dot{\phi}^0 + \bar{\eth} \dot{A}_1^0 + \frac{1}{2} \bar{\eth} \dot{B}_{11}^0 \right)= 0  \label{eq:rmbarprojection}\\
& \quad \Longrightarrow \quad -2 \sqrt{2} \bar{\eth} \phi^0 + \frac{1}{\sqrt{2}} \bar{\eth} \left(-A_1^0 + B_{11}^0 + C_{22}^0 \right) - \frac{1}{\sqrt{2}} \eth B_{22}^0 - B_{12}^0  + \bar{\eth}^{2} \left(\bar{A}_2^0 + \bar{B}_{12}^0 \right) \nonumber\\
						& \qquad \qquad  -\bar{\eth} \eth  \left(A_2^0 + B_{12}^0 \right) + \frac{1}{2} \dot{A}_{2}^1 +  \frac{1}{2}\dot{B}_{12}^1 - \sqrt{2} \bar{\eth} \left(\dot{\phi}^1 + \dot{A}_1^1 + \frac{1}{2}  \dot{B}_{11}^1 \right)= 0  \label{eq:rmbarprojection2}  \\
\bar{m}^a \bar{m}^b G_{ab}^{\prime} = 0 & \quad \Longrightarrow \quad \bar{\eth}\dot{A}_{2}^0 +  \bar{\eth} \dot{B}_{12}^0 = 0 \\
\bar{m}^a m^b G_{ab}^{\prime} = 0 & \quad \Longrightarrow \quad \dot{\phi}^0 - \frac{1}{2} \dot{B}_{11}^0 - \sqrt{2} \re \left( \eth \dot{A}_{2}^0 +  \eth \dot{B}_{12}^0 \right) + \ddot{\phi}^1 + \ddot{A}_1^1 + \frac{1}{2} \ddot{B}_{11}^1= 0 \, . \label{eq:mbarmprojection}
\end{align}
Here $u^{a}\partial_{a} = \partial/\partial u$ and $r^{a}\partial_{a} = \partial/\partial r$. As is clear from the asymptotic expansion (\ref{falloff4}) all the coefficients with superscripts $0$ and $1$ (such as $C^{0}_{22}$ and $C^{1}_{22}$) are functions of $u,\theta,\varphi$ only. Therefore, they represent limits of fields on $\scrip$. In the $G_{ab}^{\prime} u^{a}u^{b}=0,\, G_{ab}^{\prime} u^{a} \bar{m}^{b}=0$ and $G_{ab}^{\prime} r^{a}\bar{m}^{b}=0$ equations, we have provided consequences of field equations also at the sub-leading order because these equations will be directly useful. In the remaining equations we have given only the leading order consequences.

Substituting Eq.~(\ref{eq:rrprojection}) into (\ref{eq:rmbarprojection}) one finds 
\begin{equation}
\dot{A}_{2}^0 +  \dot{B}_{12}^0 = 0 \,  \label{eq:A2B12}
\end{equation}
(which in turn implies (\ref{eq:umbarprojection1})).
Using this equation in (\ref{eq:urprojection}) we find that $C_{22}^0$ is non-dynamical:
\begin{equation}
\dot{C}_{22}^0 = 0 \quad \Longrightarrow \quad C_{22}^0 \equiv C_{22}^0(\theta, \varphi) \, . \label{eq:C22}
\end{equation}
This statement holds in any gauge. In section \ref{s3.3} we will find that other equations listed above can be used to show that in fact all the leading order coefficients except $B_{22}^{0}$ are non-dynamical \emph{in the Transverse gauge} introduced in section \ref{s3.1}. 

\subsubsection{The Newman-Penrose Weyl components and Peeling}
\label{s3.2.2}

Recall that the Newman-Penrose component $\Phi^{0}_{2}(u,\theta, \varphi)$ of the asymptotic Maxwell field represents the `radiation field' at $\scrip$ and that the two radiative modes $A_{2}^{0}(u, \theta, \varphi)$ in the Maxwell connection $A_{a}$ determine $\Phi^{2}_{0}(u, \theta,\varphi)$ and $\im \Phi_{1}^{0}(u,\theta,\varphi)$ in any of the large family of gauges we considered. However, $\re \Phi_{1}^{0}$ contains `Coulombic information' that is not captured in the radiative mode $A_{2}^{0}(u,\theta,\varphi)$. 

The situation is similar in full general relativity. One thinks of the Newman-Penrose component $\Psi_{4}^{0}(u,\theta,\varphi)$ of the asymptotic Weyl tensor as representing the gravitational radiation field \cite{np}. We also know that the two radiative modes in the gravitational connection at $\scrip$ determine $\Psi_{4}^{0}(u,\theta,\varphi),\, \Psi_{3}^{0}(u,\theta,\varphi)$ and $\im\Psi_{2}^{0}(u,\theta,\varphi)$ at $\scrip$ \cite{aa-bib,aaam,aa-radmodes,aa-yau}. $\re \Psi_{2}^{0}$, on the other hand, contains `Coulombic information' that is not captured in the two radiative modes. However  these conclusions are generally arrived at using fall-off conditions that are, at least a priori, \emph{significantly stronger} than those we were led to use to accommodate the Transverse gauge in presence of sources. For example, the early literature \cite{np,etn} starts by assuming that the Newman-Penrose component $\Psi_{0}$ falls off as $1/r^{5}$, while more recent treatments work in a gauge (see, e.g., \cite{gx}) in which the conformally rescaled linearized metric has a smooth limit to $\scrip$. Nonetheless, as we will now show, this general intuition carries over even with the weaker fall-off conditions (\ref{falloff4}) because of the field equations (\ref{eq:uuprojection}) -- (\ref{eq:C22}).\\ 

The components of the Weyl tensor of interest are determined by the linearized metric coefficients as follows: 
\begin{align}
\Psi_4 &:= C_{abcd} n^a \bar{m}^b n^c \bar{m}^d = \frac{1}{r} \Psi_4^0 + \ord \left(\frac{1}{r^2} \right) \label{psi4}\\
\Psi_3 &:= C_{abcd} n^a l^b n^c \bar{m}^d =  \frac{1}{2\sqrt{2} r} \underbrace{\left(\ddot{A}_2^0 + \ddot{B}_{12}^0 \right)}_{\rm = 0 \, by \,  eq.~(\ref{eq:A2B12})} + \frac{1}{r^2} \Psi_3^0 + \ord \left(\frac{1}{r^3} \right) \label{psi3}\\
\Psi_2 &:= C_{abcd} m^a l^b n^c \bar{m}^d = 
- \frac{1}{6 r} \underbrace{\left(\ddot{\phi}^0 + \ddot{A}_1^0 + \frac{1}{2} \ddot{B}_{11}^0 \right)}_{\rm = 0 \, by \,  eq.~(\ref{eq:rrprojection})} \notag \\
& - \frac{1}{6r^2} \underbrace{\left( 3 \dot{\phi}^0 + 2 \dot{A}_1^0 + \frac{1}{2} \dot{B}_{11}^0 + \dot{C}_{22}^0  + 2 \sqrt{2} \left(\eth \dot{A}_2^0 + \eth \dot{B}_{12}^0 \right) - \sqrt{2} \left(\bar{\eth} \dot{\bar{A}}_2^0 + \bar{\eth} \dot{\bar{B}}_{12}^0 \right) + \ddot{\phi}^1 + \ddot{A}_1^1 + \frac{1}{2} \ddot{B}_{11}^1 \right)}_{\rm = 0 \, by \,  eq.~(\ref{eq:rrprojection}),(\ref{eq:mbarmprojection}),(\ref{eq:A2B12}) \, and \, (\ref{eq:C22})} \notag \\
& + \frac{1}{r^3} \left( \re \Psi_2^0 + i \, \im \Psi_2^0 \right) +  \ord \left(\frac{1}{r^4} \right)\, , \label{psi2}
\end{align}
where the leading order asymptotic fields $\Psi_{4}^{0},\, \Psi_{3}^{0}$ and $\Psi_{2}^{0}$ on $\scrip$ are given by:
\begin{align}
& \Psi_4^0 = - \ddot{B}_{22}^0  \label{psi40}\\
& \Psi_3^0 = \frac{1}{2} \bar{\eth} \dot{\phi}^0 - \frac{1}{4} \bar{\eth} \dot{B}_{11}^0 - \frac{1}{2} \eth \dot{B}_{22}^0 + \frac{1}{2} \bar{\eth} \dot{C}_{22}^0 - \frac{1}{\sqrt{2}} \dot{A}_2^0 - \frac{1}{2\sqrt{2}} \left(\ddot{A}_2^1 + \ddot{B}_{12}^1 \right) \label{psi30}\\
& \im \, \Psi_2^0 = - \frac{1}{\sqrt{2}} \im \left(2 \eth A_2^0 + \eth \dot{A}_2^1 + \eth \dot{B}_{12}^1 \right) \label{impsi20} \\
& \re \, \Psi_2^0 = -\frac{1}{2} \phi^0 - \frac{1}{4} B_{11}^0 + \frac{1}{3} \eth \bar{\eth} \left(\phi^0 + C_{22}^0 -\frac{1}{2} B_{11}^0 \right) - \frac{1}{3} \re \eth^2 B_{22}^0  - \frac{\sqrt{2}}{3} \re \eth B_{12}^0 - \frac{5}{6} \dot{\phi^1}
\notag \\
& \qquad \qquad  - \frac{1}{2} \dot{A}_1^1 - \frac{1}{12} \dot{B}_{11}^1 - \frac{1}{3} \dot{C}_{22}^1 - \frac{1}{3\sqrt{2}} \re \left( \eth \dot{A}_2^1 + \eth \dot{B}_{12}^1 \right) - \frac{1}{6}\left( \ddot{\phi}^{(2)} + \ddot{A}_1^{(2)} + \frac{1}{2} \ddot{B}_{11}^{(2)} \right)\, . \label{repsi20}
\end{align}
These equations are simply identities obtained by substituting the asymptotic expansion of $h_{ab}$ in the expression of the Weyl tensor. 

Using Einstein's equations (\ref{eq:umbarprojection2}), (\ref{eq:rrprojection}) and (\ref{eq:A2B12}), one can simplify the right side of Eq. (\ref{psi30}) to obtain:
\be \Psi_3^0 = - \eth \dot{B}_{22}^0  \label{psi302} \ee
Similarly using Eq. (\ref{eq:rmbarprojection2}) to rewrite $\eth \dot{A}_2^1 + \eth \dot{B}_{12}^1$ in $\im \Psi_2^0$ and the fact that $\phi^0, A_1^0$ etc. are real, we obtain
\begin{align}
 \im \, \Psi_2^0 &= - \im \, \eth^2 B_{22}^0 -2 \sqrt{2} \left(\eth \bar{\eth} + \frac{1}{2} \right) \im \eth \left(A_2^0 + B_{12}^0 \right) \\
& = - \im \, \eth^2 B_{22}^0 -\frac{1}{\sqrt{2}} \left(D^2 +2 \right) \im \eth \left(A_2^0 + B_{12}^0 \right)  
\label{Psi20B22}
\end{align} 
where $D^2$ is the Laplacian on the unit two-sphere.
This expression and Eq.(\ref{eq:A2B12}) now imply that $\im \dot{\Psi}_2^0$ is determined by $B_{22}^0$:
\be
 \im \, \dot{\Psi}_2^0 = - \im \, \eth^2 \dot{B}_{22}^0 \, .
\ee

Finally, field equations imply that there are relations between these three components of the Weyl tensor,
\be \dot{\Psi}_3^0 = \eth \Psi_4^0 \quad {\rm and} \quad 
 \im\,\dot{\Psi}_2^0 =  \im\,\eth \Psi_3^0\, . \label{bianchi1}
\ee
If we had used an approach to Peeling similar to the one in section \ref{s2.1} for the Maxwell field, these equations would have emerged directly from field equations on \emph{the linearized Weyl tensor}.\\

Taken together, the set of equations (\ref{psi4}) -- (\ref{bianchi1}) brings out the following important facts:
\begin{itemize}
\item Even with our weaker fall-off conditions on $h_{ab}$, field equations imply that Peeling holds for these components of the Weyl tensor:  $\Psi_{4}$ falls off as ${1}/{r}$;\, $\Psi_{3}$ as ${1}/{r^{2}}$; and $\Psi_{2}$ as ${1}/{r^{3}}$.

\item Bianchi identities inform us that apart from (important but) non-dynamical integration constants, in the linearized theory now under consideration $\Psi_{4}^{0}$ determines $\Psi_{3}^{0}$ as well as $\im \Psi_{2}^{0}$,\,\, and is itself unconstrained. Because of this property and the fact that $\Psi_{4}$ falls-off as $1/r$,\, $\Psi_{4}^{0}$ represents the \emph{radiation field} at $\scrip$. It is the analog of $\Phi_{2}^{0}$ in the Maxwell theory.

\item The leading order components $\Psi_{4}^{0}$ and $\Psi_{3}^{0}$
of the Weyl curvature are completely determined by the complex field $B_{22}^{0}$, i.e., by the part $B_{22}^{0} m_{a} m_{b} + \bar{B}_{22}^{0} \bar{m}_{a} \bar{m}_{b}$ of the leading order metric perturbation $h_{ab}$. (Interestingly, Eq. (\ref{psi40}) for $\Psi_{4}^{0}$ does not use Einstein's equations;  Eq. (\ref{psi302}) for $\Psi_{3}^{0}$, on the other hand, does.) In this sense, just as the complex field $A_{2}^{0}$ captures the two radiative modes of the Maxwell field, now the complex field $B_{22}^{0}$ represents the two radiative modes of the linearized gravitational field at $\scrip$. 

\item The component $\re\,\Psi_{2}^{0}$, on the other hand, is \emph{not} determined by $B_{22}^{0}$. Again, since $\re \Psi_{2}^{0}$ represents the `Coulombic part' of the gravitational field, this feature is consistent with our interpretation of $B_{22}^{0}$ as representing the two radiative modes of the gravitational field. 

\item In the Maxwell case, $\im \Phi_{1}^{0}$ is also completely determined by $A_{2}^{0}$ (see Eq. (\ref{Phi})). By contrast, in the gravitational case, the (first and second) leading order field equations themselves did not lead us to an analogous conclusion for $\im \Psi_{2}^{0}$: Under our weak fall-off conditions (\ref{falloff4}), we could only conclude that the \emph{time derivative} of $\im\,\dot\Psi_{2}^{0}$ is determined by $B_{22}^{0}$. Under stronger asymptotic conditions (e.g., those assumed in \cite{np}, or in the Geroch-Xanthopoulos gauge \cite{gx} used in \cite{aa-radmodes,aa-bib}), one can show that $\im \Psi_{2}^{0}$ itself is determined by $B_{22}^0$ via 
\be \im \, {\Psi}_2^0 = - \im \, \eth^2 {B}_{22}^0\, . \label{stronger} \ee
This relation in particular implies that the linearized NUT --or, the `magnetic'-- 4-momentum on $\scrip$ \cite{aaas} is zero:
\be P^{\star}_{\alpha}\, =\, \oint_{C} \alpha(\theta,\varphi)\, \im\, \Psi_{2}^{0}\, \rmd^{2}S \,= \,0 \, , \label{NUT} \ee  
where the integral is carried out over any 2-sphere cross section $C$ of $\scrip$ and $\alpha(\theta,\varphi)$ is a linear combination of the first $4$ spherical harmonics (so that $\alpha \tilde{n}^{a}$ is the restriction to $\scrip$ of a translation Killing field in Minkowski space).  

\item Interestingly, the extra term\, $-\frac{1}{\sqrt{2}} \left(D^2 +2 \right) \im \eth \left(A_2^0 + B_{12}^0 \right)$\, in Eq. (\ref{Psi20B22}) is such that this weaker relation also suffices to imply that the linearized NUT 4-momentum on $\scrip$ is zero.
\end{itemize}

\subsubsection{Radiative modes $B_{22}^{0}$ and the asymptotic shear $\sigma^{0}$}
\label{s3.2.3}

In the Newman-Penrose framework, $\Psi_{4}^{0}$ is determined by the second time derivative of $\bar\sigma^{0}$ where $\sigma^{0}$ is the asymptotic shear of the null congruence $\ell^{a}$, and $\Psi_{3}^{0}$ by the mixed angular and time derivative of $\bar\sigma^{0}$. Therefore, our relations $\Psi_{4}^{0} = -\ddot{{B}}_{22}^{0}$ and $\Psi_{3}^{0} = - \eth \dot{B}_{22}^{0}$ strongly suggest that $B_{22}^{0}$ is closely related to the asymptotic shear. However, the notion of shear requires a choice of a null congruence and a family of cross-sections. Since the framework has to be general enough to accommodate metric perturbations in the Transverse gauge, one needs to specify how this choice is made in the \emph{perturbed} space-time, assuming only the weak fall-off conditions (\ref{falloff4}) on $h_{ab}$ in place of the stronger Newman-Penrose conditions \cite{np,nt} on null tetrads and spin coefficients.  

Let us begin by making this specification. On the space-time manifold $M$, let us fix the $u,r,\theta,\varphi$ chart and the vector field $\ell^{a}$ defined by $\ell^{a}\partial_{a} = (1/\sqrt{2})\partial_{r}$ in this chart (so that it is tangential to the $u={\rm const}$ surfaces). We will consider a 1-parameter family of metrics $g_{ab}(\lambda)$ such that $g_{ab}(0) = \eta_{ab}$, and $h_{ab} = ({\rmd} g_{ab}(\lambda)/{\rmd} \lambda)|_{\lambda =0}$ and denote by $\sigma_{ab}(\lambda)$ the shear of $\ell^{a}$ on the 2-spheres $u={\rm const},\,\, r= {\rm const}$.%
\footnote{One may wish to ask that $\ell^{a}$ be null in this entire family. Then, in the linearized approximation we would have $h_{ab}\ell^{a}\ell^{b}=0$. If we were to use the transverse gauge $\Do^{a} \vh_{ab} =0$, the restricted gauge freedom, $h_{ab} \to h_{ab} + 2\nabla_{(a} f\, t_{b)}$ for any space-time function $f$, can be used to ensure that this condition is satisfied. For this choice, since $\ell^{a}$ is null w.r.t. every $g_{ab}(\lambda)$ and tangential to hypersurfaces $u={\rm const}$, it is geodetic: $\ell^{a}\,(\nabla_{a}(\lambda))\, \ell^{b} = \kappa (\lambda)\, \ell^{a}$. By an appropriate rescaling, $\ell^{a} \to \tilde\ell^{a}(\lambda) = \alpha(\lambda) \ell^{a}$ we can make it geodesic. The linearized shear $\tilde\sigma^{0}$ of $\tilde\ell^{a}$ would be the same as that of $\ell^{a}$ defined in this section.}
The shear of $\ell^{a}$ on the background Minkowski space-time vanishes. By linearized shear, we will mean:
\be \sigma_{ab} =\f{\rmd}{\rmd \lambda}\,\,\Big[\big(\nabla_{a}\ell_{b} - \frac{1}{2} S_{ab} S^{cd} \nabla_c l_d\big)\,\Big](\lambda) \Big{|}_{\lambda=0}\, \quad \hbox{\rm and we will set}\quad \sigma := m^{a}m^{b} \sigma_{ab}\, , \label{shear1}\ee
where both $\nabla_{a} \ell_{b}$ and $S_{ab}$ (the intrinsic metric on the 2-spheres $u={\rm const}, r={\rm const}$) depend on $\lambda$. We can simplify the expression of $\sigma$ considerably by using the identity $2\big(\nabla_{(a}\ell_{b)}\big)(\lambda) = {\mathcal{L}_{\ell}}\, g_{ab}(\lambda)$ for all $\lambda$,\,\,\, and the fact that, in the background Minkowski space, $m^{a}m^{b} S_{ab}=0$ and $S^{cd} \nabla_{c}\ell_{d} = \sqrt{2}/r$:
\be \sigma = \frac{1}{2} m^a m^b \Lie_\ell h_{ab} - \frac{1}{\sqrt{2}\, r} \Big(m^a m^b h_{ab}\Big) \, .\ee %\Big(\f{\sqrt{2}}{r}\Big)\ee
Finally, by substituting our asymptotic expansion (\ref{falloff4}) for the linearized metric $h_{ab}$ we obtain
\be \sigma = - \frac{1}{2 \sqrt{2}} \frac{\bar{B}_{22}^0}{r^2} + \ord \Big( \frac{1}{r^3} \Big) \quad \text{so that}\quad  \sigma^0 = - \frac{1}{2\sqrt{2}} \bar{B}_{22}^0\, . \label{shear2}\ee
Thus, as expected, the `radiative mode' $B_{22}^{0}$ at $\scrip$ is indeed the complex conjugate of the asymptotic shear $\sigma^{0}$.\\

\emph{Remark:} The Newman-Penrose framework is widely used in the literature. Therefore, it is useful to have a dictionary that spells out the relation between our conventions and those used in the Newman-Penrose framework. To be specific, for this comparison we will use the Newman-Penrose conventions from the review article by Newman and Tod in an Einstein volume \cite{nt}.

The Newman-Tod review uses a space-time metric with signature +,-,-,- while we use the signature -,+,+,+. It is simplest to keep the contravariant tetrad the same (except for factors of $\sqrt{2}$):
\be
l^a_{\rm NT} = \sqrt{2} l^a, \quad n^a_{\rm NT} = \frac{1}{\sqrt{2}} n^a, \quad m^a_{\rm NT} = m^a \quad {\rm and} \quad \bar{m}^a_{\rm NT} = \bar{m}^a 
\ee
so that the covariant vector fields are related by a minus sign:
\be
l_a^{\rm NT} = - \sqrt{2} l_a, \quad n_a^{\rm NT} = - \frac{1}{\sqrt{2}} n_a, \quad m_a^{\rm NT} = - m_a \quad {\rm and} \quad \bar{m}_a^{\rm NT} = - \bar{m}_a \, .
\ee
Since $C_{abc}{}^{d}$ does not change with signature, $C_{abcd}^{\rm NT} = - C_{abcd}$. Tetrad components of the Weyl curvature are related by
\be
\Psi_4^{\rm NT} = \frac{1}{2} \Psi_4, \quad \Psi_3^{\rm NT} = \frac{1}{\sqrt{2}} \Psi_3, \quad \Psi_2^{\rm NT} = \Psi_2
\ee
and the shear and the angular derivative operators are related by
\be \sigma^{\rm NT} = - \sqrt{2} \sigma  \quad {\rm and} \quad \eth^{\rm NT} = - \sqrt{2} \eth\, . \ee
This dictionary then recasts our expressions (\ref{psi40}) and (\ref{psi302}) of $\Psi_{4}^{0}$ and $\Psi_{3}^{0}$ and the Bianchi identities (\ref{bianchi1}) to those in \cite{nt}:
\ba \Psi_{4}^{0\, {\rm NT}} = - \ddot{\bar{\sigma}}^{0\,\rm NT} \quad
&{\rm and}& \quad \Psi_{3}^{0\, {\rm NT}} = - \eth\dot{\bar{\sigma}}^{0\,\rm NT} \nonumber\\
\dot{\Psi}_3^{0\,{\rm NT}} = - \eth^{\rm NT} \Psi_4^{0\,{\rm NT}} \quad &{\rm and}& \quad 
 \im\,\dot{\Psi}_2^{0\, {\rm NT}} =  -\im\,\eth^{{\rm NT}} \Psi_3^{0\, {\rm NT}}\, . \label{bianchi2}
\ea
%
%\bigskip

\subsection{$h_{ab}^{\TT}$ versus $h_{ab}^{\tt}$}   
\label{s3.3}

Since the discussion of section \ref{s3.2} did not refer to any specific gauge, we can use its conclusions while discussing both notions of transversality. As before we assume that in the expansion (\ref{expn}) of the linearized metric $h_{ab}$, the coefficients have the fall-off specified in (\ref{falloff4}).

\subsubsection{Transverse-Traceless modes of the linearized perturbation }
\label{s3.3.1}

Let us begin with a gauge that is well-suited to isolating the Transverse-Tracess modes. Specifically, we will assume that in addition to the fall-off conditions (\ref{falloff4}), $h_{ab}$ satisfies $\Do^{a} (\vh_{ab} - \f{1}{3} \bh \qo_{ab}) =0$,\, and $\Do^{a} \vA_{a} =0$. As in section \ref{s3.1}, these conditions fix the gauge freedom completely. Following the Maxwell case, for notational clarity, we will 
use an underbar for quantities in the Transverse-Traceless framework
and set $\ub{h}_{ab} = \vh_{ab} - \f{1}{3} \bh \qo_{ab}$.
%denote various fields of the Transverse-Traceless $\vh_{ab} - \f{1}{3} \bh \qo_{ab}$ with an underbar so that $\ub{h}_{ab}$ is Transverse-Traceless. 

Since $\ub{h}_{ab}$ is Transverse-Traceless, its Fourier transform has only two non-zero (real) components in the $\hat{k}^{a}, m^{a}(\vk), \bar{m}^{a}(\vk)$ basis in \emph{momentum space.} 
In \emph{physical space} by contrast, none of the six components of $\ub{h}_{ab}$ vanish in general. But there are relations between them.  The trace-free condition implies $\ub{B}_{11}= - 2 \ub{C}_{22}$. Other constraints among components of $\ub{h}_{ab}$ are more subtle. Let us examine the nature of these constraints to leading order in the $1/r$ expansion. As one might have anticipated from the Maxwell case, the condition $\Do^{a} \vh_{ab}=0$ implies: 
\be \partial_{u}\, \ub{B}_{11}^{0} = 0, \quad {\rm and} \quad \partial_{u}\, \ub{B}_{12}^{0} = 0 \, .\ee
Thus, in the Transverse-Traceless gauge, of the six (real) components of $\vh_{ab}$ only two, namely\, $\ub{B}_{22}^{0}$,\, have dynamical information. Furthermore none of the linearized Einstein's equations (\ref{eq:uuprojection}) -- (\ref{eq:mbarmprojection}) constrain $\ub{B}_{22}^{0}$ in any way. Thus, although the Transverse-Traceless condition does introduce restrictions on the six components of $\vh_{ab}$, the two radiative modes $\ub{B}_{22}^{0} (u,\theta,\varphi)$ are freely specifiable on $\scrip$, just as one would expect. The four other components encoded in $\ub{B}_{11}, \ub{B}_{12}$ and $\ub{C}_{22}$ of $\ub{h}_{ab}$ are non-dynamical, determined on entire $\scrip$ by their values at $i^{o}$ or $i^{+}$.

Next, let us consider the three components of $\vA_{a}$. Since $\partial_{u}\, \ub{B}_{12}^{0} =0$, Eq. (\ref{eq:A2B12}) which holds in any gauge, implies
\be \partial_{u}\, \ub{A}_{2}^{0} =0\, , \ee
and (as in the Maxwell theory) the Transversality condition $\Do^{a}{\vA}_{a}=0$ implies
\be \partial_{u}\, \ub{A}_{1}^{0} =0\, . \ee
Finally, since both $\ub{B}_{11}^{0}$ and $\ub{A}_{2}^{0}$ are non-dynamical the linearized Einstein's equation (\ref{eq:rrprojection}) implies that so is $\phic^0$:
\be \partial_{u}\, \phic^0 = 0\, . \ee 
Thus, in the Transverse-Traceless gauge, the two radiative modes $\ub{B}_{22}^{0}$ are freely specifiable on $\scrip$ but all other eight components are non-dynamical --i.e., functions only of $\theta, \varphi$-- which is the hallmark of `Coulombic information'. In terms of the $1/r$-part $\ub{h}_{ab}^{0}$ of the linearized metric, only the traceless tensor
\be \ub{B}_{22}^{0}(u,\theta,\varphi)\,\,m_{a}\,m_{b} + \bar{\ub{B}}_{22}^{0}(u,\theta,\varphi)\,\,\bar{m}_{a}\, \bar{m}_{b}\ee 
is dynamical. However, even at $\scrip$, in the Transverse-Traceless gauge $\Do^{a} \ub{h}_{ab} =0$,\, $h_{ab}^{\TT}$ is \emph{not} given by just these two terms,
\be \ub{h}_{ab}^{0}\, \not=  \, \ub{B}_{22}^{0}(u,\theta,\varphi)\,\,m_{a}\,m_{b} + \bar{\ub{B}}_{22}^{0}(u,\theta,\varphi)\,\,\bar{m}_{a}\, \bar{m}_{b}\, ; \ee
$h_{ab}^{\TT}$ also provides us with eight other non-vanishing components $\underline{\phi}^0, \ldots \ub{C}_{22}^0$ (with relations among them), as well as higher-order fields that fall-off faster than $1/r$, \emph{all of which are gauge invariant}. Therefore, using the  metric perturbation $\ub{h}_{ab}$ of the Transverse-Traceless approach we can construct fields such as $\Psi_{2}^{0}$ that carry `Coulombic information'. 

To summarize, although only two components of  $h_{ab}^{\TT}$ are dynamical at $\scrip$, it also carries other physical information through its additional components that are \emph{time-independent} on $\scrip$. Thus, the overall structure is completely analogous to that in the Maxwell case.

\subsubsection{Transverse projection in the physical space}
\label{s3.3.2}

As explained in section \ref{s1}, the second notion of transversality is local in physical space: One simply projects the full linearized metric $h_{ab}$ into the 2-spheres $u= {\rm const}, \, r= {\rm const}$ in space-time using the projection operator $P_{a}^{b} = m_{a}\bar{m}^{b} + \bar{m}_{a} m^{b}$. The desired $h_{ab}^{\tt}$ is then obtained by removing the trace from $P_{a}^{c} P_{b}^{d} h_{cd}$. Thus the $tt$ projection corresponds simply to \emph{discarding 8 of the 10 components of $h_{ab}$ in space-time.}

Clearly the tt-projection is not gauge invariant since $h_{ab}^{\tt}$ can change in an uncontrolled fashion under $h_{ab} \to h_{ab} + 2 \nabla_{(a} \xi_{b)}$ where $\xi^{a}$ is an arbitrary vector field subject only to the condition that the gauge transformation should preserve the class of linearized metric perturbations that satisfy (\ref{falloff4}). Indeed, even the leading order, $1/r$-part of $h_{ab}^{\tt}$ fails to be invariant. However, as in the Maxwell case, one can impose the Lorenz gauge $\nabla^{a}h_{ab} =0$ and also require that $h_{ab}$ be traceless in a neighborhood of $\scrip$. Then the gauge freedom is restricted to:
\be h_{ab} \to h_{ab} + 2 \nabla_{(a} \xi_{b)}\quad {\rm where} \quad 
\Box\, \xi^{a} =0 \,\,\, {\rm and} \,\,\, \nabla^{a} \xi_{a} =0 \ee
in that neighborhood. As in the Maxwell case, even under these additional gauge conditions, some of the leading order coefficients in $h_{ab}$ fail to be gauge invariant: For example $\phi^{0} \to \phi^{0} + \partial_{u} \xi^{0}_{u}$ and $A_{1}^{0} \to A_{1}^{0} + \partial_{u} (\xi^{0}_{r} - \xi^{0}_{u})$. However, the leading order, $1/r$-part of $h_{ab}^{\tt}$ is now gauge invariant. We can decompose it using the basis vectors $m_{a}, \bar{m}_{a}$ on 2-spheres $u={\rm const}, r={\rm const}$:
\be h_{ab}^{\tt} = \ut{B}_{22}\,m_{a}m_{b} + \bar{\ut{B}}_{22}\, \bar{m}_{a} \bar{m}_{b}\, , \ee
where, as in the Maxwell case, the under-tilde is a reminder that we are referring to metric coefficients in the tt-approach. From now on, we will assume that $h_{ab}$ of the tt approach is in the Lorenz and traceless gauge in a neighborhood of $\scrip$.

To summarize, in the tt-approach we have access only to the two components $\ut{B}_{22}$ of the metric perturbations and only the leading order asymptotic part $\ut{B}_{22}^{0}$ of this field is gauge invariant. Field equations (\ref{eq:uuprojection}) -- (\ref{eq:mbarmprojection}) do not constrain these leading order components in any way. They suffice to determine $\Psi_{4}^{0},\,\Psi_{3}^{0}$ and $\im \dot\Psi_{2}^{0}$ completely but 
carry no information about the `Coulombic aspect' of the solution registered in, e.g., $\re \Psi_{2}^{0}$. Again the situation is completely analogous to that in the Maxwell case.

\subsubsection{Comparison}
\label{s3.3.3}

We can now compare the two approaches. Because the linearized metric $h_{ab}$ is fixed completely in the Transverse-Traceless gauge, all the underbarred metric coefficients --not just the leading order ones, carrying the superscript $0$-- are fully gauge invariant. In the tt-approach on the other hand only the two components encoded in $\ut{B}_{22}^{0}$ of the \emph{leading order} metric perturbation are gauge invariant. Now, components $\Psi_{4}^{0}, \ldots \Psi_{0}^{0}$ of the linearized Weyl curvature are manifestly gauge invariant and for the entire class of perturbations we considered, $\Psi_{4}^{0}$ is determined by the second time derivative of the coefficient $B_{22}^{0}$. Therefore, in particular, we know that the radiative modes in the two approaches are related by:
\be \Psi_{4}^{0} = - \ddot{\ub{B}}_{22}^{0} = - \ddot{\ut{B}}_{22}^{0} \, . \ee
Similarly $\Psi_{3}^{0}$ is gauge invariant and from Eq. (\ref{psi302}) we conclude:
\be \Psi_{3}^{0} = -\eth\, \dot{\ub{B}}_{22}^{0} = -\eth\, \dot{\ut{B}}_{22}^{0}\ee
Since $\ub{B}_{22}^{0}$ and $\ut{B}_{22}^{0}$ have spin weight $-2$ on which $\eth$ only has a trivial kernel, these two equations imply 
\be \dot{\ub{B}}_{22}^{0} = \dot{\ut{B}}_{22}^{0}\,. \ee
From the relation (\ref{shear2}) between the asymptotic shear $\sigma^{0}$ and $B_{22}^{0}$ that holds in any gauge, and the fact that the time derivative of the shear determines the Bondi news tensor  \cite{aa-radmodes,aa-bib}, we conclude that the (linearized) Bondi news tensor $N_{ab}^{0}$ is given by
\be N_{ab}^{0} = \f{1}{\sqrt{2}}\,\, \Big(\dot{B}^{0}_{22}\,\, \bar{m}_{a}\, \bar{m}_{b} + \dot{\bar{B}}^{0}_{22}\,\, {m}_{a} \, {m}_{b} \Big)\, ,    \ee
where $B_{22}^{0}$ can be evaluated using the TT or the tt method. Similarly, $\im\, \dot{\Psi}_{2}^{0} = - \im\,\eth^{2} \dot{B}_{22}^{0}$ can be evaluated using \emph{either} $\ub{B}_{22}^{0}$ \emph{or} $\ut{B}_{22}^{0}$.

Note however, that there is no a priori guarantee that the two sets of radiative modes would be equal. In general, we only know that 
\be \ub{B}_{22}^{0} - \ut{B}_{22}^{0} = f(\theta,\varphi) 
\label{difference} \ee
for some complex-valued field $f$ of spin weight $-2$. Thus, although the leading order fields $\ub{B}_{22}^{0}$ and $\ut{B}_{22}^{0}$ are both gauge invariant, in general they will differ by a non-dynamical $f(\theta,\varphi)$. Eq.(\ref{shear2}) implies that the asymptotic shears
${\underline\sigma}^{0}$ and ${\ut\sigma}^{0}$ computed in the two approaches will also differ. In the Maxwell case, we could conclude that the analogous spin weight $-1$ field $f$ is purely electric. For linearized gravity, we did not find such a restriction (because our weaker fall-off conditions (\ref{falloff4}) did not allow us to conclude that (\ref{stronger}) must hold). 

Can this difference (\ref{difference}) have an observable effect? A gravitational wave detector measures strain that is directly caused by the metric perturbation. The strain $h_{ab}^{\tt}$ predicted by the tt approach differs from that predicted by the TT approach even to leading asymptotic order: At $\scrip$ we have
\be h_{ab}^{tt}\big|_{\scrip} \equiv \ut{B}_{22}^{0}\,\,m_{a}\,m_{b}\,+\, \bar{\ut{B}}_{22}^{0}\,\, \bar{m}_{a} \, \bar{m}_{b}\, \not= \, \ub{B}_{22}^{0}\,\,m_{a}\,m_{b}\, + \, \bar{\ub{B}}_{22}^{0}\,\, \bar{m}_{a} \, \bar{m}_{b}  \ee
However, the two predictions for the wave form will be shifted simply by a \emph{non-dynamical} term given by $f$ in (\ref{difference}) and therefore will be unobservable.

\subsubsection{Soft charges, gravitational memory and fluxes across $\scrip$}
\label{s3.3.4}

`Soft' charges associated with gravitons label infrared sectors at null infinity which first emerged in the setting of asymptotic quantization of the non-linear gravitational field \cite{aa-radmodes,aa-asymquantization,aa-bib,aa-yau}. While the initial setting was non-perturbative, rooted in techniques developed in algebraic quantum field theory, they play 
an important role also in perturbative treatments of graviton scattering that take into account subtleties associated with infrared issues (see, e.g. \cite{jwrsrk}). There has been a resurgence of interest in these soft charges in connection with Weinberg's theorems on soft gravitons in perturbative quantum gravity (see, e.g. \cite{softtheorem,mcal}). However, as in the Maxwell case, the definition and properties of gravitational soft charges can be discussed entirely in the classical framework.

Given any non-dynamical test function $\tilde\alpha(\theta,\varphi)$ on $\scrip$, 
there is a soft graviton charge $Q_{\tilde\alpha}$ defined by integrating the Bondi news $N^{0}(u,\theta,\varphi) := N^{0}_{ab}m^{a}m^{b}$ against $\tilde{\alpha}(\theta,\varphi)$:
\be Q_{\tilde\alpha} = \int_{\scrip} \tilde\alpha(\theta,\varphi) N^0 (u,\theta,\varphi)\, \rmd u\, \rmd^{2}S = \f{1}{\sqrt{2}}\,\,\Big(\oint_{u=\infty} - \oint_{u=-\infty}\Big)\, \tilde{\alpha} \, B_{22}^{0}\, \rmd^{2}S.\ee
Since the TT and tt approach define the same Bondi News $N^0$ --or, equivalently, since $\ub{B}_{22}^{0} - \ut{B}_{22}^{0}$ is $u$-independent-- the two approaches define the same soft charges.%
\footnote{In place of the infinitely many $Q_{\tilde\alpha}$, one for each $\tilde\alpha(\theta,\varphi)$, one sometimes defines a charge $Q(\theta,\phi)$ obtained by integrating $N^0$ along each generator of $\scrip$. These charges are complex-valued because $N^0$ is complex-valued. However, under stronger fall-off conditions that are generally used in the literature, the asymptotic shear can be taken to be `purely electric' and $Q(\theta,\phi)$ reduces to a real-valued function.}\\

Next, let us consider gravitational memory in the linearized framework  \cite{memory3}. It encodes the displacement that the detector is left with after the passage of a pulse of gravitational wave. As in the Maxwell case, it can be expressed as a special case of the soft charge $Q_{\tilde\alpha}$ where the test field $\tilde\alpha$ is a Dirac delta distribution peaked at the generator of $\scrip$ singled out by the location of the detector. Therefore, again, the memory calculated in the two frameworks agrees.\\

Finally let us consider fluxes of energy-momentum and angular momentum carried by gravitational waves across $\scrip$. In the Maxwell theory, the task was straightforward because of the availability of the stress-energy tensor. In the gravitational case, there is no gauge invariant definition of stress energy tensor even in the asymptotic region. Therefore one is forced to choose another strategy. In the case of energy-momentum, there is a well established expression of the flux carried away by gravitational waves in full non-linear general relativity, obtained using two different considerations: (i) appealing to the asymptotic field equations (see, e.g., \cite{bondi-sachs,np,rg}),  and, (ii) calculating Hamiltonians generating canonical transformations \cite{aaam,aams}, corresponding to the Bondi-Metzner-Sachs translations on $\scrip$ \cite{sachs}. Therefore, for linearized gravity, we can obtain the desired expression of flux starting with the full theory  (by performing a second order linearization of the full flux). Then, in the notation used in section \ref{s2.3.3}, translation Killing fields $K^{a}$ have the form $K^{a} \= \alpha(\theta,\phi) \tilde{n}^{a}$ at $\scrip$, where $\alpha(\theta,\phi)$ is a linear combination of the first $4$ spherical harmonics. The flux of the component of linear momentum along $K^{a}$ is given by 
\ba \mathcal{F}_{\alpha} &=& \f{1}{32\pi G} \int_{\scrip} \alpha(\theta,\phi) \, N_{ab}N_{cd}\, \hat\eta^{ac} \hat\eta^{bd}\,\, \rmd u\, \rmd^{2}S \nonumber\\
&=& \f{1}{32\pi G} \int_{\scrip} \alpha(\theta,\phi) \, |\dot{B}_{22}^{0}|^{2}
\,\, \rmd u \,\rmd^{2}S\, . \ea
Since $\dot{\ub{B}}_{22}^{0} = \dot{\ut{B}}_{22}^{0}$, the two approaches yield the same energy-momentum flux across $\scrip$.\\

For angular momentum, the situation is less clearcut. First of all, in the full theory there is a `supertranslation ambiguity' because the Bondi-Metzner-Sachs group does not admit a canonical Poincar\'e group \cite{sachs}. But this ambiguity disappears in the linearized theory because now the background geometry provides us with a natural Poincar\'e group. But a second issue remains. Unlike in the case of energy-momentum, field equations have not directly yielded a viable expression of the flux of (or 2-sphere `charge' integrals for) angular momentum. The definition of the flux that is agreed upon in the literature \cite{tdms,wz} arises, again, as the Hamiltonian generating canonical transformations on the radiative phase space, induced by rotations and boosts \cite{aams}. It involves not only the Bondi news but also the asymptotic shear at $\scrip$. If we were to start with this expression in the full theory and carry out the reduction tailored to the linear theory, we would conclude that the TT and the tt approach would yield different fluxes because $\underline\sigma^{0}$ differs from $\ut\sigma^{0}$ in general. The situation would be similar to that in the Maxwell case. 

However, there is a further caveat. The phase space of radiative modes on $\scrip$ was obtained starting from the covariant phase space solutions to \emph{vacuum} Einstein equations (in full general relativity) \cite{aaam}. To understand the associated subtleties, it is instructive to revisit the Maxwell theory. There, one can also construct the phase space of radiative modes starting from source-free solutions to Maxwell equations, and obtain expressions of Hamiltonians generating canonical transformations corresponding to translations and rotations \cite{aams}. For source-free solutions, these Hamiltonians yield the same expressions of fluxes of energy-momentum and angular momentum carried away by electromagnetic waves across $\scrip$ as those obtained using the stress-energy tensor. If we add sources and consider retarded solutions, the agreement continues for energy-momentum. However, for angular momentum, this Hamiltonian method does not capture the `Coulombic term' involving $G(\theta,\varphi)$ in (\ref{amflux2})! In retrospect this is not surprising because the phase space knows only about the radiative degrees of freedom. Very recently, we reanalyzed the covariant phase space of Maxwell theory, allowing for sources and using retarded solutions. We found that the presence of sources \emph{introduces a surface term} in the expression of the symplectic structure --i.e. 2-sphere integrals at $i^{o}$ and $i^{+}$, in addition to a 3-surface integral over all of $\scrip$-- that encapsulates the additional `Coulombic information' in the retarded solution. Thus, the previous radiative phase space has to be appropriately augmented to allow for sources. When this is done, the expression for the flux of energy-momentum is unaltered, but that for angular momentum now includes the additional term involving $G(\theta,\varphi)$ so that the phase space flux expression agrees with the correct one, obtained using the stress-energy tensor. We believe that the situation will be similar for linearized (and full) general relativity. But that analysis has only begun.\\

To summarize, in general, the two notions of transversality of metric perturbations lead to different fields $\ub{B}^{0}_{22}$ and $\ut{B}_{22}^{0}$ on $\scrip$ representing the two radiative modes. However, this difference disappears in calculations of soft charges, gravitational wave memory and flux of energy-momentum carried by gravitational waves across $\scrip$. There is a strong indication that, as in the Maxwell case, only the TT approach will provide us with the correct expression of the flux of angular momentum in presence of matter sources. However, to fully establish this result, one would have to extend the existing phase space of gravitational `radiative modes' to allow for the presence of sources, or devise another reliable method to calculate fluxes across $\scrip$.\\

\emph{Remark:} As noted in section \ref{s1}, the difference between the TT decomposition and the tt projection was already pointed out in \cite{racz1}. However, the subsequent discussion in \cite{racz1} focused on corrections to linearized gravity due to `back reaction' of an effective stress-energy tensor associated with gravitational waves, and possible significance of these corrections to laser interferometric gravitational wave detectors. By contrast, in this paper we worked strictly within linearized gravity and addressed two quite different issues: (i) Why the deep conceptual differences between the TT decomposition and the tt projection have not been noticed in the main stream literature for so long; (ii) What the precise relation between them is; and, \, (iii) How one would characterize physical quantities for which the difference is important.

\section{Discussion}
\label{s4}

It is commonplace in monographs and review papers on gravitational waves to begin with the technically simpler setting of electromagnetic waves because the conceptual issues and the necessary mathematical methods are rather similar. In the Maxwell theory, one can forego vector potentials $A_{a}$ altogether and express all physical observables directly in terms of the field $F_{ab}$. But in linearized gravity, we cannot express even the basic physical quantities such as the flux of energy, momentum and angular momentum directly in terms of linearized curvature; we have to work with potentials. Therefore, as a prelude to gravity, it is customary to first cast Maxwell theory also in terms of potentials $A_{a}$ and extract the two physical degrees of freedom in electromagnetic waves by considering their transverse parts. However, in this literature one finds \emph{two} notions of transversality. In the first, one requires the spatial projection $\vA_{a}$ of $A_{a}$ to satisfy $\Do^{a} \vA_{a} =0$. This condition becomes algebraic in momentum space enabling one to express the Transverse vector potential $\vA_{a}^{\,\,\T}$ as
\be \vA_{a}^{\,\, \T}(t,\vx) = \f{1}{(2\pi)^{\f{3}{2}}}\,\, \int \big(\alpha_{2}(t, \vk)\, m_{a}(\vk) + \bar{\alpha}_{2}(t, \vk)\, \bar{m}_{a}(\vk)\big)\, e^{i\vk\cdot\vx}\, \rmd^{3} k  \label{FT} \ee
for some functions $\alpha_{2}(t, \vk)$, where $m_{a}(\vk), \, \bar{m}_{a}(\vk)$ are (complex) null basis vectors in the \emph{momentum space}, transverse to the radial vector field $k^{a}$. The second notion, that of $\vA_{a}^{\,\,\t}$, is local in physical space:
\be \vA_{a}^{\,\, \t}(t, \vx) = \ut{A}_{2}(t, \vx)\, m_{a}(\vx) + \ut{\bar{A}}_{2}(t, \vx)\, \bar{m}_{a}(\vx) \ee
where $m_{a}(\vx),  \bar{m}_{a}(\vx)$ are now (complex) null vector fields in the \emph{physical space} transverse to the radial vector field $r^{a}$. Because $m^{a}(\vk)$ is \emph{not} a constant vector field in the momentum space, there is no relation between the two notions {except when $\alpha_2 (t, \vk) = \alpha_{0}(t)\, \delta^{3}(\vk, \vk_{0})$ for some $\vk_{0}$, i.e., when $\vA_{a}(t, \vx)$ is a plane wave, \emph{and} one focuses attention only on the 2-plane in the physical space, orthogonal to the propagation direction $\vk_{0}$.} Yet, often one finds that the two notions are simply identified, perhaps because of the rough intuition that far away from the source the field resembles a plane wave. However, one then expresses physical quantities such as the energy, momentum, and angular momentum, carried away by electromagnetic waves in terms of these vector potentials --quantities which are infinite for plane waves! In any case, what is of direct physical interest are appropriate superpositions of plane waves as in (\ref{FT}), and not plane waves themselves. For these superpositions $\vA_{a}^{\,\T}$ and $\vA_{a}^{\,\,\t}$ are entirely different fields.

In section \ref{s2} we showed that null infinity $\scrip$ serves as a useful platform to compare and contrast the two notions since it is the natural arena to study electromagnetic (and gravitational) radiation. However we could not directly use the rich machinery available in the literature because much of it is tailored to gauges in which the 4-vector potential $A_{a}$ admits a smooth limit to $\scrip$ while in general $A_{a}$ in the Transverse gauge does not. Therefore we had to introduce a more general framework in which the vector potential is allowed to satisfy weaker fall-off conditions. Although the analysis becomes a bit more complicated, in the end we found that, for retarded solutions, many of the familiar relations between fields and potentials continue to hold at $\scrip$ even with these weaker fall-off conditions. 

Interestingly, even though the two notions of transversality are so very different, we found that the two frameworks lead to closely related notions of radiative modes at $\scrip$. They are captured in freely specifiable, gauge invariant complex fields with spin weight $-1$, which we denoted by $\ub{A}_{2}^{0}(u,\theta,\phi)$ in the  $A_{a}^{\T}$ framework and by $\ut{A}_{2}^{0} (u,\theta,\phi)$ in the $A_{a}^{\t}$ framework. The radiative modes at $\scrip$ suffice to determine a number of quantities of physical interest. In any given physical situation, generically $\ub{A}_{2}^{0} \not= \ut{A}_{2}^{0}$. But their difference is characterized by a \emph{non-dynamical} field $f(\theta,\phi)$ of spin weight $-1$ (which is purely electric). We showed that this fact has important implications: predictions of the two frameworks are identical for the electromagnetic memory, soft charges and (local and global) flux of energy-momentum carried by electromagnetic waves across $\scrip$. \emph{These results shed considerable light on why the basic conceptual difference between the two notions has been overlooked so often.}

However, the $A_{a}^{\T}$ framework provides additional \emph{gauge invariant} fields at $\scrip$ and these carry the `Coulombic information' in the solution, enabling us to calculate, for example, the total charge. The projected $A_{a}^{\t}$ does \emph{not} have this information: The only gauge invariant fields this framework provides at $\scrip$ are the two radiative modes, $\ut{A}_{2}^{0}$. An unforeseen finding was that this difference plays a crucial role in the expression of the (local and global) flux of angular momentum across $\scrip$: In presence of sources carrying non-zero electric charge, this flux can be calculated only in the $A_{a}^{\T}$ framework since its expression also contains fields other than the radiative modes. In the $A_{a}^{\t}$ framework one may be tempted to retain just those terms that involve the radiative modes $\ut{A}_{2}^{0}$. But then one would get the wrong answer: For Maxwell fields in Minkowski space-time we know what the correct flux is using the stress energy tensor and Killing fields. As explained in section \ref{s2.3.3}, this is a subtle issue that arises only in presence of sources: For source-free electromagnetic waves, angular momentum flux \emph{can be} expressed entirely in terms of the two radiative modes and the two frameworks yield the same answer, while in presence of sources, fields representing the `Coulombic aspect' of the solution also enter the flux expression. Finally, note that this phenomenon arises for Maxwell fields in Minkowski space-times where, as we have noted, the notion of angular momentum is completely unambiguous. Thus, the \emph{phenomenon is unrelated to supertranslation ambiguities and possible failure of Peeling properties.}

In section \ref{s3}, we discussed linearized gravitational fields. Again, there are two notions, $h_{ab}^{\TT}$ and $h_{ab}^{\tt}$, of transverse-traceless modes that are often used interchangeably in the literature: Imposing $\Do^{a} \big({\vh}_{ab} - \f{1}{3} (\qo^{cd}\vh_{cd}) \qo_{ab}\big)= 0$ versus using the projection $(P_a^b\,P_c^d - \f{1}{2} P_{ab}P^{cd}) \vh_{ab}$ into the $r={\rm const}$, $u={\rm const}$ 2-spheres. 
We called the first `Transverse-Traceless' gauge condition,  and the second,  `transverse-traceless' projection. For exactly the same reasons as we discussed above in the Maxwell case, the two notions are unrelated, except for plane waves. Again, we could not directly use results available in the literature (e.g., \cite{np,nt,aa-radmodes,aams}) because they are obtained in gauges (e.g. the one given in \cite{gx}) in which the conformally rescaled metric perturbation admits a smooth limit to $\scrip$, while in gauges of interest to our discussion, it does not. Therefore we introduced weaker fall-off conditions motivated by solutions to field equations in the Transverse gauge. Because of this generality, the discussion becomes quite cumbersome at a technical level but it turns out that many --though not all-- of standard results in the literature, derived using stronger fall-off conditions, continue to hold also in gauges we considered. Calculations are technically much more complicated than those in the Maxwell case, but the overall structure is very similar. Therefore findings of section \ref{s2} provided guidance to organize our discussion.

As in the Maxwell case, each of the two approaches provides us with complex fields on $\scrip$ (with spin weight $-2$) that represent the two radiative modes of the linearized gravitational field. We denote them by $\ub{B}_{22}^{0}$ in Transverse-Traceless gauge, and by $\ut{B}_{22}^{0}$ in the transverse-traceless projection approach. Both functions are unconstrained by field equations (as well as by asymptotic conditions) but in general $\ub{B}_{22}^{0} \not= \ut{B}_{22}^{0}$. However, again, the difference is encoded in a non-dynamical field $f(\theta,\varphi)$ on $\scrip$. As a consequence, one can calculate the Bondi News tensor, the flux of energy-momentum carried across $\scrip$, soft charges and gravitational memory using either notion and obtain the same results. 

Thus, our discussion provides considerable reassurance: It shows that although the two approaches to metric perturbations are very different conceptually, dropping the distinction does not introduce errors in many results of  `practical importance'. The underlying reason is that these results depend only on the time derivatives of the radiative modes --the Bondi news-- and are therefore insensitive to the `Coulombic information' in the solution. However, for results that \emph{are} sensitive to the `Coulombic information', as in the Maxwell case, the two frameworks are quite different: the only gauge invariant fields that the transverse-traceless projection provides are the radiative modes $\ut{B}_{22}^{0}$, while the Transverse-Traceless gauge provides us with a host of other gauge invariant fields from which we can, for example, obtain $\re \Psi_{2}^{0}$ that determines the Bondi 4-momentum in the linear theory. As explained in section \ref{s3.3.4}, we expect that this difference will be important for the expression of the flux of angular momentum across $\scrip$, just as it is in the Maxwell case. However, because we do not have a gauge invariant stress energy tensor for the gravitational field, further work is needed to settle this issue definitively.\\

We will conclude with three remarks.
\begin{itemize}

\item This analysis in the linearized context has already brought out 
the fact that there is a subtle but important difference between properties of source-free solutions to Maxwell or Einstein's equations and those with sources, thereby opening an unforeseen window. As we explained in section \ref{s3.3.4} in the Maxwell case it has led to a natural extension of the phase space of radiative modes at $\scrip$ that also incorporates the asymptotic `Coulombic properties' of solutions through an additional surface term to the symplectic structure, without having to enlarge the phase space to  include the degrees of freedom corresponding to sources. It also suggests a similar extension for both linearized and full general relativity. To construct the extended framework in detail is a challenging task but likely to provide even qualitatively new insights.

\item The primary motivation for this paper came from gravitational waves produced by isolated sources such as the coalescence of compact bodies. For these systems both notions of transversality are available in the linear approximation, but in practice one generally uses the transverse projection. Indeed, in the gravitational wave community that develops approximation methods there is often some unease about using the Transverse-Traceless gauge because the decomposition into Transverse and Longitudinal parts is non-local in space. Since the interaction of waves with the detector is local, one asks, how could it involve a Transverse field that requires global considerations? But recall that for a ring of test particles, the tidal deformation is governed by the (linearized) Riemann tensor that is locally defined and, as explained in standard books, it can be expressed as the second time derivative of the metric perturbation $h_{ab}^{\TT}$ in the Transverse-Traceless gauge (see, e.g. \cite{mtw,rmw}). The physical detector interacts with gauge invariant quantities and because $h_{ab}^{\TT}$ \emph{is} gauge invariant everywhere (not just asymptotically) one could even say that the detector interacts locally with these degrees of freedom. In this context it is useful to note again that in the transverse-projection approach, only the radiative modes $\ut{B}_{22}^{0}$ extracted from the $1/r$-part of the perturbation are gauge invariant.

\item In the cosmological context, on the other hand, we do not have asymptotic flatness. Therefore, we neither have access to $\ut{B}_{22}^{0}$, nor another way to extract gauge invariant radiative modes using the transverse projection. That is why the cosmology literature uses only the Transverse-Traceless approach.%
\footnote{In the community that focuses on gravitational waves from isolated systems, there is sometimes an unease on whether  the TT decomposition is well-defined in the cosmological context: How can the Fourier transforms needed for this decomposition be well-defined when we have a spatially homogeneous matter distribution? Indeed, it is well-defined only after the homogeneous part of the perturbation is absorbed into the background and the `purely inhomogeneous modes' are sufficiently regular, say square integrable. (Note that if they are square-integrable for some choice of origin in the physical space, they are so for any other choice.)}
 However, so far the focus of this literature has been on primordial gravitational waves that are source-free solutions to linearized Einstein's equations rather than waves produced by compact sources. As gravitational wave observatories extend their reach, especially through space based missions, they will open a new window at the interface of astrophysics and cosmology. To make full use of the ensuing opportunities, one would need theoretical tools to analyze gravitational waves from astrophysical sources at cosmological distances. The current idealization presupposes that the source and detector are both in an asymptotically flat space-time and uses $\scrip$ or, equivalently, $1/r$-expansions. To fully exploit the cosmological potential of the next generation of gravitational wave observatories, it would be appropriate to start developing a framework that takes us beyond the pristine, 50 year old paradigm developed by Bondi, Sachs, Newman, Penrose and others \cite{bondi-sachs,np,nt}. In particular, at the linearized level this requires us to go beyond the transverse projection method. This has been possible for isolated systems in de Sitter --rather than Minkowski-- space-times where $\scrip$ is space-like rather than null, and new approximation methods rooted in the Transverse gauge are both necessary and viable  \cite{abk2,abk-prl,abk3}. There has also been some progress in extending the fully non-linear Bondi-Sachs framework to the asymptotically de Sitter context \cite{aa-ropp}. The next major step would be to generalize that framework to Friedmann-Lema\^{i}tre-Robertson-Walker backgrounds in presence of a positive cosmological constant.

\end{itemize}

\section*{Acknowledgment} We thank Badri Krishnan, Eric Poisson, Istv\'{a}n R\'{a}cz and especially Aruna Kesavan for discussions. This work was supported in part by the NSF grant  PHY-1505411, the Eberly research funds of Penn State and Mebus Graduate Fellowship to BB. AA thanks the Erwin Schr\"odinger Institute in Vienna for hospitality during preparation of this manuscript.

\appendix

\section{Linearized Einstein's equations in the Transverse gauge}
\label{a1}

In this Appendix we provide some details of how the linearized Einstein's equations used in section \ref{s3} are obtained. As in the main text, we perform a space-time decomposition of the linearized metric $h_{ab}$,  
\be 2\phi = t^{a}t^{b} h_{ab};\quad \vA_{a} = \qo_{a}^{b} t^{c} h_{bc}; \quad {\rm and} \quad \vh_{ab} = \qo_{a}^{c}\, \qo_{b}^{d} h_{cd} \ee
and of the (linearized) stress-energy tensor $T_{ab}$
\be \label{source}
\rho := t^a t^b T_{ab}, \quad \vec{J}_a :=  \qo_a^{\, b} \, t^c  \, T_{bc}, \quad  \vec{T}_{ab} := \qo_a^{\, m} \qo_b^{\,n} T_{mn}\, .
\ee
Finally, since $T_{ab}$ is smooth and of compact spatial support, we can further decompose it into longitudinal and transverse parts:  $\vec{J}_{a} =  \vec{J}^{\,\,T}_{a} + \vec{J}^{\,\, \L}_{a}$ and $\vec{T}_{ab} =  \vec{T}^{\,\,T}_{ab} + \vec{T}^{\,\, \L}_{ab}$.

Recall that linearization of Einstein's equations off Minkowski space, without imposing any gauge condition, yields:
\be \label{EE} 
-\Box h_{ab} + 2 \nabla_{(a} \nabla^{c} h_{b)c} - \nabla_{a} \nabla_{b} h + \big(\Box h - \nabla^{c}\nabla^{d} h_{cd}\big)\, \eta_{ab} = 16\pi G \, T_{ab} \,
\ee
where $h= \eta^{ab}h_{ab}$ is the 4-trace of $h_{ab}$. To begin with, let us impose the Transverse gauge condition only on $\vh_{ab}$: $\Do^{a}\vh_{ab} =0$. (The restricted gauge freedom will be removed later.) Then by a space-time decomposition, the linearized Einstein's equations reduce to:
\begin{align} 
& \Do^{2} \bh =  -16 \pi G \, \rho\quad {\rm where}\quad \bh = \qo^{ab}\vh_{ab} %&\qquad {\rm (time-time)} 
\label{timetime}\\
& \Do^{2}\vec{A}_a^{\,\T} 
= - 16 \pi G \vec{J}_a^{\,\T} \equiv -16\pi G \vec{J}_a - \Do_{a} (\mathcal{L}_{t}\bh)  %&\qquad {\rm (space-time)} 
\label{spacetime}\\
& \Box \vec{h}_{ab} \, +\, 2 \Do_{(a} \dot{\vA}_{b)}\,+\, \Do_{a} \Do_{b} (\bh -2 \phi)\,+\, \big(2\Do^{2}\phi - 2 \Do^{c}\dot{\vA}_{c} - \Box \hb \big) \qo_{ab}\,   =\, - 16 \pi G \, \vec{T}_{ab} \label{spacespace1} %&\,\, {\rm (space-space)} 
\end{align}

\emph{Remarks:} 

1. It is obvious from Eq. (\ref{timetime}) that we would have had a conflict with linearized Einstein's equations if we had demanded, in addition to the Transversality condition $\Do^{a}h_{ab} =0$, that the 3-trace $\bh$ of $\vh_{ab}$ should vanish. 

2. In Eq. (\ref{spacetime}) we have decomposed $\vA_{a}$ into its Longitudinal and Transverse parts. Since in section \ref{s3.1} we begin by asking only that $h_{ab}$ be smooth and vanish at spatial infinity, this decomposition is to be understood in the distributional sense. However, in the end, properties of $T_{ab}$ and the linearized field equations ensure that $\vA_{a}^{\,\,\T}$ is in fact a suitably differentiable tensor field in space-time.\\

Let us simplify Eq.~(\ref{spacespace1}), the space-space projection. By taking its trace and using (\ref{timetime}), we find that the  term proportional to $\qo_{ab}$ in (\ref{spacespace1}) can be expressed in terms of sources:
\be 2\Do^{2}\phi - 2 \Do^{c}\dot{\vA}_{c} - \Box \hb  = - 8 \pi G (\bar{T} -\rho) \ee
where $\bar{T} = \qo^{ab} \vec{T}_{ab}$. Substituting this term in (\ref{spacespace1}) one obtains
\be \Box \vh_{ab} + \Do_{a} \Do_{b} \bh + 2 \big(\Do_{(a} \dot{\vA}_{b)} - \Do_{a} \Do_{b} \phi) = - 16\pi G \tilde{T}_{ab}\quad {\rm where} \quad \tilde{T}_{ab} := \vec{T}_{ab} + \f{1}{2}\big(\rho - \bar{T}\big)\qo_{ab} \, . \ee
Next, let us make a Transverse and Longitudinal decomposition of this equation to obtain:
\ba & \Box \vh_{ab} \, \equiv\, \Box \vh_{ab}^{\,\, \T}\, = \, -16\pi G \tilde{T}_{ab}^{\,\T} \label{spacespace2} \\
& \Do_{a} \Do_{b} (\bh - 2 \phi) + 2 \Do_{(a} \dot{\vA}_{b)} = -16\pi G \tilde{T}_{ab}^{\,\L} \label{spacespace3} \ea

To simplify these equations further, we fix the remaining gauge freedom by requiring, as in section \ref{s3.1}, that $\vA_{a}$ be Transverse, i.e., satisfy $\Do^{a}\vA_{a} =0$, and $\phi$ satisfy the  condition $\phi \to 0$ as $r\to \infty$ for all $t$. Then, using (\ref{timetime}) again as well as conservation of the stress-energy tensor, (\ref{spacespace3}) yields:
\be \Do^{2}\phi  = - 4 \pi G \big(\rho + \bar{T} - 2 \dot{J}^{L} \big)  \label{spacespace4}\ee
where $J^{\L}$ is the potential for $\vec{J}^{\,\, \L}_{a}$:\,\, $\Do_{a}\, J^{\L}\, =\, \vec{J}^{\,\,\L}_{a}$. Thus, in the fully fixed Transverse framework, dynamics of the metric components $\phi, \vA_{a}^{\,\, \T}$ and $\vh_{ab}^{\,\, \T}$ is governed by Eqs. (\ref{timetime}),\, (\ref{spacetime}) and  (\ref{spacespace4}). These are the equations we used in section \ref{s3.1}. \\

\emph{Remark:} Since the trace of (\ref{spacespace2}) yields a hyperbolic equation for $\bh$ and (\ref{timetime}) provides us with an elliptic equation also for $\bh$, it is natural to inquire if they are consistent. Algebraic simplification boils this question down to the consistency between $\Do^{2} \ddot{\bh} = -16\pi G\, \Do^{2}\dot{J}^{\L}$ and $\Do^{2} \bh = -16\pi G \rho$, which is ensured by conservation of the stress energy tensor.


\begin{thebibliography}{99}

\bibitem{Einstein:1916}A.~Einstein, N\"aherungsweise Integration der Feldgleichungen der Gravitation,  Sitzungsberichte der K\"{o}niglich Preussischen Akademie der Wissenschaften, Berlin, (1916). 

\bibitem{ligo} 
  B.~P.~Abbott {\it et al.} [LIGO Scientific and Virgo Collaborations],
  Observation of Gravitational Waves from a Binary Black Hole Merger,
  Phys.\ Rev.\ Lett.\  {\bf 116}, no. 6, 061102 (2016)
%  doi:10.1103/PhysRevLett.116.061102
%  [arXiv:1602.03837 [gr-qc]].

\bibitem{primordial} 
	K.~N.~Abazajian {\it et al.} [CMB-S4 Collaboration], CMB-S4 Science Book, First Edition, \texttt{arXiv:1610.02743}.

\bibitem{pw} E.~Poisson and C.~Will, \emph{Gravity: Newtonian, Post-Newtonian, Relativistic} (Cambridge University Press, Cambridge 2005), Chapters 5.5 and 11. 

\bibitem{straumann}N.~Straumann, \emph{General Relativity with Applications to Astrophysics} (Springer, Berlin, 2004), Chapter 4.

\bibitem{mtw}C.~W.~Misner, K.~S.~Thorne and J.~A.~Wheeler, \emph{Gravitation} (W.~H.~Freeman, San Fransisco, 1973), Chapters 35-37. 

\bibitem{hughes} S.~A.~Hughes, Probing strong-field gravity and black holes with gravitational waves, Proceedings of the 19th Japanese Workshop on General Relativity and Gravitation, \texttt{arXiv:1002:2591}.

\bibitem{bondi-sachs} H.~Bondi, M.~van der Burg, and A.~Metzner, Gravitational waves in genera1 relativity VII. Waves from axi-symmetric isolated systems, Proc. R. Soc. (London) A\textbf{269}, 21 (1962);\\
R.~K.~Sachs, Gravitational waves in general relativity VIII. Waves in asymptotically flat space-times Proc. R. Soc. (London) A \textbf{270}, 103 (1962).

\bibitem{rpwr}R.~Penrose and W.~Rindler, \emph{Spinors and Space-time: Volume 1, Two-Spinor Calculus and Relativistic Fields} (Cambridge University Press, Cambridge, 1984), Section 9.7. 

\bibitem{ahm}A.~Ashtekar, G.~T.~Horowitz and A.~Magnon, A generalized tensor calculus and its applications to physics, Gen. Rel. Grav. \textbf{14}, 411-428 (1982). 

\bibitem{aabb1} A.~Ashtekar and B.~Bonga, On a basic conceptual confusion in gravitational radiation theory, Class. Quant. Grav. \textbf{34}, 20LT01 (2017).

\bibitem{racz1} I. R\'{a}cz, Gravitational radiation and isotropic change of the spatial geometry, \texttt{arXiv:0912.0128}.

\bibitem{racz2} A.~Frenkel and I.~R\'{a}cz, On the use of projection operators in electrodynamics, Eur. J. Phys. \textbf{36} 015022 (2015).

\bibitem{tdms} T.~Dray and M.~Streubel, Angular momentum at null infinity, Class. Quant. Grav. \textbf{1}, 15-26 (1984).

\bibitem{ak-thesis} A.~Kesavan, Asymptotic structure of space-time with a positive cosmological constant, Ph.D. Dissertation, The Pennsylvania State University, (2016), Section 2.4.1 and Chapters 3 and 5.
            
\bibitem{np} R.~Penrose and E.~T.~Newman, New conservation laws for zero rest mass fields in asymptotically flat space-times, Proc. R. Soc. (London) A\textbf{305}, 175-204 (1968);\\ 
Approach to gravitational radiation by a method of spin coefficients, J. Math. Phys. \textbf{3}, 566-578 (1962).

\bibitem{etn} E.~T.~Newman, private communication to AA (2017).

\bibitem{dcsk}D.~Christodoulou and S.~Klainerman, \textit{The
    global non-linear stability of Minkowski space} (PUP, Princeton,
    1993).

\bibitem{asgr21} A.~Strominger, How to grow hair on a black hole, Plenary talk at the GR21 conference, New York, July 14th 2016.

\bibitem{pced} P.~T.~Chrusciel and E.~Delay, Existence of non-trivial, vacuum, asymptotically simple space-times, Class. Quant. Grav. \textbf{19}  L71-L80 (2002). 

\bibitem{aa-bib} A.~Ashtekar, \textit{Asymptotic Quantization}
    (Bibliopolis, Naples, 1987); available at 
    http://igpg.gravity.psu.edu/research/asymquant-book.pdf
   
\bibitem{aa-yau} A.~Ashtekar, Geometry and physics of null infinity, prepared for Surveys in Differential Geometry, edited by L. Bieri and S. T.- Yau, pp99-122, (International press, Boston, 2015 ); \texttt{arXiv: 1409:1800}.
      
\bibitem{ep} E.~Poisson, private communication to AA (2016).    

\bibitem{fk} P.~P.~Kulish and L.~D.~Faddeev, Asymptotic conditions
    and infrared divergences in quantum electrodynamics, Teor. Mat.
    Fiz. \textbf{4} 153-170 (1970).
    
\bibitem{aakn} A.~Ashtekar and K.~S.~Narain, Infrared problems in quantum field theory and Penrose's null infinity, Syracuse University Report, Presented at the VIth International Conference on
Mathematical Physics (1981). 

\bibitem{krakow} A.~Herdegen, Asymptotic structure of electrodynamics revisited, Lett. Math.Phys. \textbf{107}, 1439-1470 (2017)    

\bibitem{bg} L.~Bieri and D.~Garfinkle, An electromagnetic analogue of gravitational wave memory, Class.\ Quant.\ Grav.\  {\bf 30}, 195009 (2013).

\bibitem{gx} R.~Geroch and B.~C.~Xanthopoulos, Asymptotic simplicity is stable, \textbf{19}, 714-718 (1978).

\bibitem{aa-radmodes} A.~Ashtekar, Radiative Degrees of Freedom of the Gravitational Field in Exact General Relativity, Journal of Mathematical Physics, 22, 2885-2895 (1981).  

\bibitem{aams} A.~Ashtekar and M.~Streubel, Symplectic geometry of
radiative modes and conserved quantities at null infinity, Proc.
R. Soc. (London) \textbf{A376}, 585-607 (1981).
 
\bibitem{aaam} A.~Ashtekar and A.~Magnon, On the symplectic
    structure of general relativity,  Comm. Math. Phys. \textbf{86},
    55-68 (1982). 

\bibitem{aa-radmodes} A.~Ashtekar, Radiative Degrees of Freedom of the Gravitational Field in Exact General Relativity, Journal of Mathematical Physics, 22, 2885-2895 (1981). 

\bibitem{aaas} A.~Ashtekar and A.~Sen, NUT 4-momenta are forever, J. Math. Phys. \textbf{23}, 2168-2178 (1982).
    
\bibitem{nt} E.~T.~Newman and K.~P.~Todd, Asymptotically flat space-times in \emph{General Relativity and Gravitation 2}, edited by A.~Held (Plenum, New York, 1980).    
 
\bibitem{rg}R.~Geroch, In: \emph{Asymptotic structure of
    space-time}, Edited by L. Witten, pp1-106 (Plenum, New
    York, 1976).
  
\bibitem{aa-asymquantization} A.~Ashtekar, Asymptotic Quantization of the Gravitational Field, Physical Review Letters, \textbf{46}, 
573-577 (1981).

\bibitem{jwrsrk} J.~Ware, R.~Saotome and R.~Akhouri, Construction of an asymptotic S-matrix for perturbative gravity, JHEP \textbf{10}, 159 (2013).
 
\bibitem{softtheorem} T.~He, V.~Lysov, P.~Mitra, and A.~Strominger,
     BMS supertranslations and Weinberg's soft graviton theorem,
     JHEP {\bf 1505}, 151 (2015).
  
\bibitem{mcal} M.~Campiglia and A.~Laddha,
	Asymptotic symmetries and subleading soft graviton theorem,
  	Phys.\ Rev.\ D {\bf 90}, no. 12, 124028 (2014).
  
\bibitem{memory3} 
	Ya. B. Zel'dovich and A. G. Polnarev, Radiation of gravitational waves by a cluster of superdense stars,	Sov. Astron. \textbf{18}, 17 (1974);\\
	L. Bieri and D. Garfinkle, A perturbative and gauge
    invariant treatment of gravitational wave memory, Phys. Rev. 
    D\textbf{89}, 084039 (2014).
    
\bibitem{sachs} R.~Sachs, Asymptotic symmetries in gravitation
    theory, Phys. Rev. D\textbf{128}, 2851-2864 (1962).    
 
\bibitem{wz}R.~M.~Wald and A.~Zoupas, A definition of `conserved quantities' in general relativity and other theories of gravity, Phys.Rev. D\textbf{61}, 084027 (2000).
   
\bibitem{rmw}R.~M.~Wald, \emph{General Relativity} (The University of Chicago Press, Chicago, 1984), Chapter 4.  

\bibitem{abk2} A.~Ashtekar, B.~Bonga and A.~Kesavan, Asymptotics with a positive cosmological constant: II.  Linear fields on de Sitter space-time, Phys. Rev. D \textbf{92}, 044011 (2015).

\bibitem{abk-prl}A.~Ashtekar, B.~Bonga and A.~Kesavan, Gravitational waves from isolated systems: Surprising consequences of a positive cosmological constant, Physical Review Letters \textbf{116}, 051101 (2016).

\bibitem{abk3}A.~Ashtekar, B.~Bonga and A.~Kesavan, Asymptotics with a positive cosmological constant:III. The quadrupole formula, Phys. Rev. D\textbf{92}, 10432 (2015).

\bibitem{aa-ropp} A.~Ashtekar, Implications of a positive cosmological constant for General Relativity, Rep. Prog. Phys. \textbf{80}, 102901 (2017)

\end{thebibliography}
\end{document}